\documentclass[aps,prl,reprint,superscriptaddress,longbibliography,floatfix]{revtex4-2}

\usepackage[T1]{fontenc}
\usepackage{amsmath,amssymb,bm}
\usepackage{graphicx,booktabs}
\usepackage[colorlinks=true,linkcolor=blue,citecolor=blue,urlcolor=blue]{hyperref}
\newcommand{\Tr}{\operatorname{Tr}}
\newcommand{\E}{\mathbb E}

\newcommand{\lettersection}[1]{\par\addvspace{0.35\baselineskip}\noindent\textbf{\textit{#1.---}}\ignorespaces}
\usepackage{needspace}

\newtheorem{smlemma}{Lemma}[section]
\newtheorem{smtheorem}[smlemma]{Theorem}
\newenvironment{smproof}{\par\noindent\textit{Proof.}\ \ignorespaces}{\unskip\nobreak\hfill\(\square\)\par\medskip}

\usepackage{etoolbox}
\makeatletter
\newcommand{\OmitLetterContents}{\let\l@section\@gobbletwo}
\newcommand{\ReferencePart}{main}
\AtBeginDocument{%
  \patchcmd{\endthebibliography}{\label{LastBibItem}}%
    {\label{LastBibItem-\ReferencePart}}%
    {}{\PackageError{main}{Bibliography label patch failed}{Check the REVTeX version.}}%
  \expandafter\let\expandafter\r@LastBibItem\csname r@LastBibItem-main\endcsname
  \let\auto@bib\@empty
}
\newcommand{\StartSupplementReferences}{%
  \renewcommand{\ReferencePart}{supp}%
  \expandafter\let\expandafter\r@LastBibItem\csname r@LastBibItem-supp\endcsname
}
\makeatother

\begin{document}

\title{Emergent Recursive Kernels in Quantum Reservoir Computing}
\author{Ce Wang}
\date{\today}

\begin{abstract}
Learning with a quantum reservoir can be far simpler than simulating its dynamics. We show that a quantum reservoir governed by a fixed evolution, consisting of an input coupling followed by an independently sampled Haar unitary, admits an exact recursive kernel in the large-dimension limit at fixed history length. Memory orthogonality and factorization of propagated overlaps reduce each kernel evaluation to a scalar recurrence, linear in history length and independent of reservoir size. We establish self-averaging around this limit and connect the kernel coefficients to out-of-time-ordered correlators. Our analysis of finite-dimensional effects shows that the computational simplicity of the Haar kernel does not come at the expense of expressive power. Moreover, we demonstrate how the kernel can be tuned for different learning tasks by adjusting properties such as memory retention and temporal interaction order. On the NARMA10 benchmark, the tuned kernel achieves lower prediction error than the tested Ising and Sachdev--Ye--Kitaev reservoirs.
\end{abstract}
\maketitle

\lettersection{Introduction}Quantum reservoir computing (QRC) is an active field that combines quantum physics with machine learning~\cite{Fujii2017}. It harnesses rich quantum dynamics and high-dimensional feature spaces to encode input histories for time-series prediction and nonlinear information processing. Recent studies have advanced our understanding of QRC through perspectives including the edge of chaos~\cite{Martinez2021,Kobayashi2025,CaoXiaLi2026}, information-processing capacity~\cite{Dambre2012,Martinez2020,Benavoli2026KIPC}, and quantum resources such as entanglement and magic~\cite{Kora2024EntanglementQRC,Xia2026MagicQRC}. Experimental progress in nuclear-spin, superconducting, and photonic platforms has also demonstrated temporal information processing and controllable memory~\cite{Negoro2018NMR,Suzuki2022Natural,Yasuda2023RepeatedMeasurements,Paparelle2026OpticalQRC}.

A defining feature of QRC is that the reservoir dynamics remains fixed, while training typically adjusts only the weights of a linear readout. From a machine-learning perspective, QRC with a linear readout trained by ridge regression admits an equivalent kernel ridge regression formulation~\cite{GrossRieser2026}. Quantum dynamics maps input histories to features, whose inner products define the associated temporal kernel~\cite{Schuld2021Kernels}. Analytical temporal kernels, including recurrent and Volterra constructions, have been developed for classical reservoirs~\cite{Hermans2012,Dong2020,Gonon2022,Malacarne2026}. Related kernel formulations for QRC use overlaps between reservoir states or kernels constructed from measured reservoir features~\cite{GrossRieser2026,Aaraba2026QuaRK}.

An important lesson of kernel methods is that even an infinite-dimensional feature space can admit an efficiently computable kernel, with well-known examples including Gaussian kernels and infinite-width neural tangent kernels~\cite{Jacot2018NTK}. Thus, the exponential growth of a quantum feature space does not by itself establish a quantum advantage in learning~\cite{Xiong2025}. Studies of information-processing capacity and expressivity in QRC have likewise highlighted constraints imposed by input encoding and accessible readouts~\cite{Martinez2020,Schuette2025Expressivity}. This motivates us to explore whether the temporal kernels induced by QRC admit a simple analytical description.

Here we identify a mechanism by which an efficiently computable recursive kernel emerges in QRC. For a fixed writer and an independently sampled Haar unitary reused throughout the input stream, memory orthogonality and propagation factorization yield a closed scalar recurrence in the large-dimension limit at fixed history length. Interestingly, we find a connection between the kernel coefficients and out-of-time-ordered correlators (OTOCs). We further show that, for local writers, finite-dimensional deviations of the exact kernel from its recursive approximation do not preclude efficient classical prediction. Finally, we use this recursive structure to construct a classical kernel family with interpretable memory controls. On the NARMA10 benchmark, this family achieves lower mean prediction error than commonly studied Hamiltonian-driven QRC models, such as Ising and Sachdev--Ye--Kitaev (SYK) reservoirs, over the system sizes and training budgets studied.

\begin{figure*}[t]
 \centering
 \includegraphics[width=\textwidth]{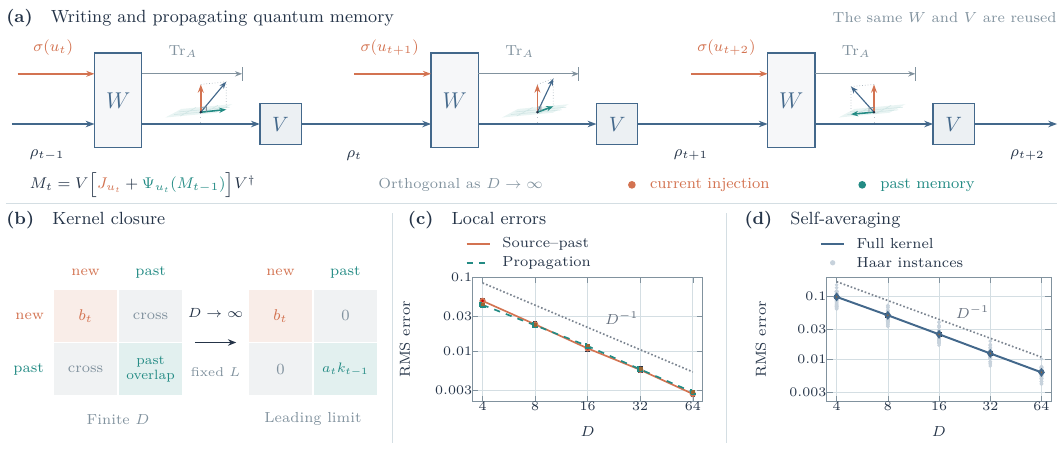}
 \caption{Orthogonal memory and a self-averaging recursive kernel. (a) Fixed writer $W$ and Haar router $V$, with injection orthogonal to propagated memory as $D\to\infty$ at fixed history length. (b) Cross-term suppression and factorized propagation yield the recurrence. (c) Root-mean-square (RMS) source--memory overlaps and propagation residuals, using the exact $k_{t-1}$. (d) RMS kernel difference $k_L-\kappa_L$; faint points show individual-router errors. Panels (c,d) use all pairs of 128 length-$16$ histories initialized at $I/D$, and 48 routers per $D$. Dotted lines indicate $D^{-1}$ scaling; bars give 95\% bootstrap intervals across routers.}
 \label{fig:protocol}
\end{figure*}

\lettersection{Model and exact kernel}We consider the standard repeated-input QRC protocol~\cite{Fujii2017}. An input $u\in[0,1]$ prepares a fresh qubit $A$ in $\sigma(u)=|\chi(u)\rangle\langle\chi(u)|$, with $|\chi(u)\rangle=\sqrt{1-u}|0\rangle+\sqrt u|1\rangle$. A fixed joint unitary $U$ couples it to a retained system of dimension $D$, after which $A$ is discarded:
\begin{equation}
 \rho_t=\Tr_A[U(\sigma(u_t)\otimes\rho_{t-1})U^\dagger].
 \label{eq:protocol}
\end{equation}
For input histories $\boldsymbol u=(u_1,\ldots,u_L)$ and $\boldsymbol v=(v_1,\ldots,v_L)$, let $\rho_t$ and $\rho'_t$ denote the states after their first $t$ inputs. We consider fitting a linear readout to a complete set of features $\Tr(O_\alpha\rho_t)$, such as all nonidentity Pauli expectations, with traceless Hermitian $O_\alpha$ normalized by $\Tr(O_\alpha O_\beta)=D\delta_{\alpha\beta}$. Ridge regression on these features is equivalent to kernel ridge regression~\cite{GrossRieser2026} with
\begin{align}
 k_t&=\sum_\alpha\Tr(O_\alpha\rho_t)\,\Tr(O_\alpha\rho'_t)\nonumber\\
 &=D\Tr(\rho_t\rho'_t)-1=\frac{\Tr(M_tM'_t)}D,
 \label{eq:readout-kernel}
\end{align}
where $M_t=D\rho_t-I$ and $M'_t=D\rho'_t-I$ represent the accumulated memories. This normalization assigns each feature, such as an individual Pauli expectation value, a dimension-independent weight in the kernel.

To reveal how a recursive kernel emerges from quantum dynamics, we consider the construction $U=(I_A\otimes V)W$ illustrated in Fig.~\ref{fig:protocol}(a). Here $W$ couples the input qubit to the reservoir, while $V$ is an independently sampled Haar-random unitary acting only on the reservoir. The memory then evolves as
\begin{equation}
 M_t=V[J_{u_t}+\Psi_{u_t}(M_{t-1})]V^\dagger.
 \label{eq:signal-update}
\end{equation}
Here $\Psi_u$ is the quantum channel determined by the current input,
\begin{align}
 \Psi_u(O)&=\sum_{\mu=0}^{1}K_\mu(u)\,O\,K_\mu(u)^\dagger,\nonumber\\
 K_\mu(u)&=(\langle\mu|_A\otimes I)\,W\,(|\chi(u)\rangle_A\otimes I),
 \label{eq:writer-channel}
\end{align}
where $\mu=0,1$ labels the computational basis of the discarded output qubit, and $J_u=\Psi_u(I)-I$ is the injected response.

\lettersection{Recursive kernel in the Haar limit}Memory orthogonality is central to the emergence of the recursive kernel~\cite{White2004,WangQiu2026}. Here, the Haar-random evolution scrambles previously injected memories, suppressing their overlap with the directions of the current injection even after propagation through the input-dependent channel. We consider histories initialized at the maximally mixed state $I/D$, so that memory is built entirely from the injected inputs. With $u=u_t$ and $v=v_t$, the kernel expands as
\begin{align}
 k_t={}&\frac{\Tr(J_uJ_v)}D
       +\frac{\Tr[J_u\Psi_v(M'_{t-1})]}D\nonumber\\
 &+\frac{\Tr[\Psi_u(M_{t-1})J_v]}D
       +\frac{\Tr[\Psi_u(M_{t-1})\Psi_v(M'_{t-1})]}D.
 \label{eq:overlap-expansion}
\end{align}
At large $D$ and fixed history length, the cross terms satisfy
\begin{align}
 \frac{\Tr[J_u\Psi_v(M'_{t-1})]}D&=O(D^{-1}),\nonumber\\
 \frac{\Tr[\Psi_u(M_{t-1})J_v]}D&=O(D^{-1}),
 \label{eq:source-past}
\end{align}
expressing the near-orthogonality between current injection and propagated memory illustrated in Fig.~\ref{fig:protocol}(a). The $O(D^{-1})$ terms denote root-mean-square errors over Haar realizations. Mathematically, Eq.~\eqref{eq:source-past} follows from asymptotic freeness, with finite-dimensional corrections controlled by Haar moment estimates~\cite{SM}. The same tools simplify the overlap of two propagated memories to
\begin{equation}
 \frac{\Tr[\Psi_u(M_{t-1})\Psi_v(M'_{t-1})]}D
 =a(u,v)k_{t-1}+O(D^{-1}),
 \label{eq:memory-propagation}
\end{equation}
where $k_{t-1}=\Tr(M_{t-1}M'_{t-1})/D$ is the preceding kernel and
\begin{equation}
 a(u,v)=\sum_{\mu,\nu=0}^{1}
 \left|\frac{\Tr[K_\mu(u)^\dagger K_\nu(v)]}D\right|^2.
 \label{eq:local}
\end{equation}
Thus, propagation preserves the previous memory overlap up to a scalar factor determined by the current inputs. Defining the source kernel
\begin{equation}
 b(u,v)=\frac{\Tr(J_uJ_v)}D,
 \label{eq:source-kernel}
\end{equation}
and writing $a_t=a(u_t,v_t)$ and $b_t=b(u_t,v_t)$, we obtain $k_t=b_t+a_tk_{t-1}+O(D^{-1})$. Combining these reductions yields our main result, a closed recursive temporal kernel generated by the Haar-limit dynamics, as illustrated in Fig.~\ref{fig:protocol}(b),
\begin{equation}
 \boxed{\kappa_t=b_t+a_t\kappa_{t-1},\qquad\kappa_0=0,}
 \label{eq:recursion}
\end{equation}
with the explicit form
\begin{equation}
 \kappa_L(\boldsymbol u,\boldsymbol v)=\sum_{j=1}^{L}b_j\prod_{s=j+1}^{L}a_s.
 \label{eq:age}
\end{equation}
Since $a$ and $b$ are themselves positive-semidefinite kernels, the recurrence defines a valid temporal kernel for every history length that can be used directly for learning. With the writer coefficients held fixed, it is the exact temporal kernel in the Haar limit~\cite{SM}.

The connection to finite-dimensional reservoirs is quantified by self-averaging. For fixed history length $L$, Haar integration and concentration give~\cite{CollinsSniady2006,CollinsMale2014,MeckesMeckes2013,SM}
\begin{equation}
 (\E_V|k_L-\kappa_L|^2)^{1/2}=O(D^{-1}).
 \label{eq:convergence}
\end{equation}
Thus, individual reservoir kernels concentrate around this recursive kernel as $D$ increases, with the predicted scaling observed in the local reductions [Fig.~\ref{fig:protocol}(c)] and the full-kernel error [Fig.~\ref{fig:protocol}(d)]. The estimates retain the local coefficients at the actual dimension $D$; their constants may depend on $L$~\cite{SM}.

The recursive kernel can be evaluated directly from the input histories in $O(L)$ scalar operations per pair once its local coefficients are known, independently of the reservoir dimension. It therefore provides a learning model whose evaluation requires no quantum-state preparation, propagation, or measurement. For comparison, obtaining the feature kernel of a finite reservoir through conventional dense simulation costs $O(LD^3)$. Recent constructions offer efficient quantum routes to Haar statistics~\cite{Ma2025RandomUnitaries,Schuster2025LowDepth}. Even without accounting for the cost of constructing and implementing the quantum circuits, standard independent SWAP readout requires $O(LD^2)$ reservoir updates for repeated state preparation at fixed additive kernel accuracy and confidence~\cite{Ekert2002DirectEstimation,SM}.

We now make three further remarks to clarify the structure and physical interpretation of the recursive kernel.

\noindent\textit{Connection to OTOCs.} Define $\Delta_u=2\sigma(u)-I_A$, with input-site operators implicitly tensored with $I_D$. We derive the identity~\cite{SM}
\begin{equation}
 b(u,v)=\frac14\sum_{P_A\in\{I_A,X_A,Y_A,Z_A\}}
 \frac{\Tr[\Delta_u U^\dagger P_AU\,\Delta_v U^\dagger P_AU]}{2D}.
 \label{eq:one-step-otoc}
\end{equation}
For Hamiltonian evolution $U=e^{-iH\Delta t}$, $U^\dagger P_AU$ is the Heisenberg-evolved Pauli operator. Equation~\eqref{eq:one-step-otoc} therefore expresses $b(u,v)$ as a Pauli average of infinite-temperature OTOCs for equal inputs, and their mixed-input generalization for unequal inputs. This establishes a direct analytical connection between OTOCs and the kernel underlying QRC learning, complementing earlier numerical observations linking OTOC decay to reservoir memory performance~\cite{Kutvonen2020}. The identity is exact at finite $D$ and does not rely on Haar randomness. For the same input-injection protocol and an arbitrary joint unitary $U$, the coefficients of the exact temporal kernel can be expressed as sums of multi-time operator correlations, including higher-order OTOCs~\cite{SM}.

\noindent\textit{Factorization freedom.} For any reservoir unitary $Q$, the transformation $V\mapsto VQ$, $W\mapsto(I_A\otimes Q^\dagger)W$ preserves the full evolution $U=(I_A\otimes V)W$ and the local kernels $a,b$. This freedom raises an apparent paradox. Starting from a local update $U=W_{\mathrm{loc}}$ with an identity router, any Haar-sampled $V$ yields the factorization $U=(I_A\otimes V)\widetilde W$, with $\widetilde W=(I_A\otimes V^\dagger)W_{\mathrm{loc}}$. The Haar-limit result might therefore seem applicable even to dynamics confined to a fixed local reservoir block. However, this construction enforces an exact dependence between the writer and router, whereas our theory fixes the writer \emph{before} independently sampling the Haar router. Once $\widetilde W$ is held fixed, reproducing the original local evolution requires a uniquely matched router, which constitutes a zero-measure event under independent Haar sampling~\cite{SM}.

\noindent\textit{Memory--injection tradeoff.} For local kernels generated by the unitary QRC update considered here, the diagonal values $a(u,u)$ and $b(u,u)$ quantify memory retention and injection strength, respectively, and satisfy~\cite{SM}
\[
 b(u,u)+2a(u,u)\le2.
\]
Starting from $\kappa_0=0$, induction gives $\kappa_t\le b_t+2a_t\le2$ on the diagonal; positive semidefiniteness then yields $|\kappa_L(\boldsymbol u,\boldsymbol v)|\le2$ for arbitrary histories.

\begin{figure}[!t]
 \centering
 \includegraphics[width=\columnwidth]{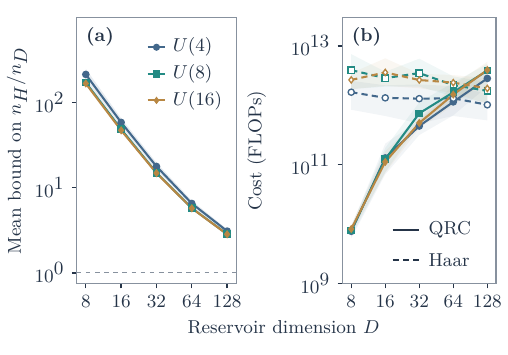}
 \caption{Finite-dimensional learning gains and classical costs. (a) Device-averaged upper bound on the sufficient sample-budget ratio $n_H/n_D$ [Eq.~\eqref{eq:sample-certificate}], for $U(4)$, $U(8)$, and $U(16)$ writers. (b) Total modeled operations for direct QRC simulation (solid) and the Haar kernel (dashed), at measured training sizes reaching the same test MSE. Both use full-kernel eigendecomposition; costs include representation construction, fitting, and validation. Points average twelve writers and four routers each; shading gives pointwise 95\% hierarchical-bootstrap intervals.}
 \label{fig:finite}
\end{figure}

\lettersection{Finite-dimensional effects and learning costs}We next show that the computational simplicity of the Haar kernel does not come at the expense of expressive power. For finite histories initialized at the maximally mixed state, finite-dimensional dynamics recombines and reweights the corresponding Haar recursive features, so every linear reservoir readout remains representable~\cite{SM}. This inclusion does not, however, guarantee comparable learning performance on every task. If the same target requires larger readout weights with the Haar kernel, it is more weakly represented by its features and can require more data to learn reliably. To quantify this effect, let $n_D,n_H$ denote the sufficient sample budgets supplied by a common generalization bound at the same target expected test MSE~\cite{BartlettMendelson2002,SM}. For any target readable by the finite-dimensional reservoir,
\begin{equation}
 \frac{n_H}{n_D}\le
 \lambda_{\max}\!\left(\mathbf K_k^{1/2}\mathbf K_\kappa^{+}\mathbf K_k^{1/2}\right).
 \label{eq:sample-certificate}
\end{equation}
Here, $\mathbf K_k$ and $\mathbf K_\kappa$ are the corresponding kernel matrices on a common complete probe set, following the same normalization procedure; $\lambda_{\max}$ denotes the largest eigenvalue and $+$ the pseudoinverse. The maximizing direction identifies the target most favorable to the finite-dimensional representation under this bound~\cite{Huang2021Power,Kornjaca2024,Kar2026PhotonicKernel}.

\begin{figure}[!t]
 \centering
 \includegraphics[width=\columnwidth]{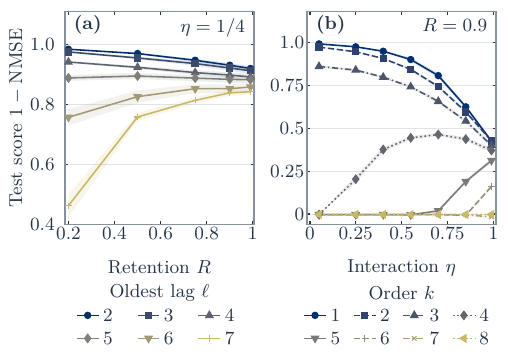}
 \caption{Memory allocation and temporal interaction order. Independent length-eight histories use $z\sim\mathrm U[-1,1]$, with lag zero current. (a) Consecutive products $z_{\ell-2}z_{\ell-1}z_\ell$, $\ell=2,\ldots,7$, at $\eta=1/4$. (b) Products $\prod_{j\in\mathcal A}z_j$, $7\in\mathcal A\subseteq\{0,\ldots,7\}$, grouped by $k=|\mathcal A|$, at $R=0.9$. Lines and bands show means and pointwise 95\% confidence intervals over eight datasets with 512 training examples. Targets have unit variance; training and validation label-noise variance is $0.1$, and test targets are clean.}
 \label{fig:design}
\end{figure}

To examine how the potential learning advantage of finite-dimensional effects changes with reservoir size, we vary $D$ while keeping the corresponding Haar kernel fixed. We randomly sample local writers $w\in U(2d)$, with $d=2,4,8$, and embed each as $W_D=w\otimes I_{D/d}$. For each device, we choose the target that maximizes the theoretical sample-budget ratio. For all three choices of $d$, this maximal ratio exhibits a similar rapid decline with $D$, with its average over devices falling from approximately $170$--$220$ at $D=8$ to about $3$ at $D=128$ [Fig.~\ref{fig:finite}(a)]. Thus, the strongest potential learning gains from finite-dimensional effects are concentrated in small reservoirs, where they can be retained through inexpensive direct classical simulation.

Actually, the measured ratios of mean training sizes needed to reach the same test MSE are well below the theoretical budget ratios in Fig.~\ref{fig:finite}(a). To account for both kernel evaluation costs and data requirements, we compare the total estimated operation counts for representation construction, fitting, and validation, with both models using eigendecomposition of their full training kernel matrices [Fig.~\ref{fig:finite}(b)]. Direct simulation is cheaper at small $D$, whereas the cost ordering reverses as the reservoir grows and the two models' data requirements become closer. Thus, even for targets maximizing the finite reservoir's theoretical sample advantage, direct simulation provides an inexpensive route at small $D$, while the Haar kernel reaches the same accuracy with comparable data requirements at lower computational cost for larger reservoirs.

\lettersection{Interpretable kernel controls}The scalar recurrence separates input injection from the retention and subsequent modulation of earlier features. For the single-qubit encoding, define $z(u)=2u-1$, $h(u)=\sqrt{1-z(u)^2}$, and $\boldsymbol s(u)=(h(u),z(u))^{\mathsf T}$. The local kernels generated by a unitary writer take the form
\begin{align}
 b(u,v)&=\boldsymbol s(u)^{\mathsf T}B\,\boldsymbol s(v),\nonumber\\
 a(u,v)&=a_0+\boldsymbol s(u)^{\mathsf T}C\,\boldsymbol s(v),
 \label{eq:seven}
\end{align}
\begin{samepage}
where the real symmetric matrices $B,C$, together with $a_0$, specify seven coefficients jointly constrained by the writer.
\par\end{samepage}

\begin{samepage}
To demonstrate the tunability of the recursive kernel, we consider a simple two-parameter classical family without requiring a unitary realization:
\begin{align}
 b(u,v)&=\frac{1-R}{2}\bigl[h(u)h(v)+z(u)z(v)\bigr],\nonumber\\
 a(u,v)&=R\bigl\{(1-\eta)+\eta\bigl[h(u)h(v)+z(u)z(v)\bigr]\bigr\},
 \label{eq:design}
\end{align}
where $0<R<1$ and $0\le\eta\le1$.
\par\end{samepage}

The resulting flexibility in learning can be seen through tasks that vary in memory depth and interaction order. Contributions injected $\ell$ steps earlier carry a factor $R^\ell$, so increasing $R$ shifts relative weight toward older inputs. This redistribution becomes visible when a product of three consecutive inputs is shifted across delays: at $\eta=1/4$, larger $R$ improves the most distant targets at the expense of recent ones [Fig.~\ref{fig:design}(a)]. The parameter $\eta$, in turn, balances constant and input-dependent propagation, changing how earlier features combine with subsequent inputs. To probe this effect, we fix $R=0.9$ and compare products containing the oldest input, grouped by their number of factors. Keeping the required memory depth fixed reveals a preference for different interaction orders: low-order products favor smaller $\eta$, whereas fifth- and sixth-order products improve at larger $\eta$ [Fig.~\ref{fig:design}(b)].

\lettersection{Comparison with Hamiltonian reservoirs}We next compare the recursive kernel with Ising and complex Sachdev--Ye--Kitaev reservoirs on NARMA10 using continuous input streams~\cite{Paquot2012,Kobayashi2025}. Each model undergoes an initial washout and retains its memory across the training, validation, and test periods.

The physical baselines are randomly coupled Ising and complex SYK reservoirs~\cite{Fujii2017,Martinez2021,Kobayashi2025}, described by
\begin{align}
 H_{\mathrm{Ising}}&=\sum_{i<j}J_{ij}X_iX_j+h_z\sum_i Z_i,\nonumber\\
 H_{\mathrm{SYK}}&=H_4+\alpha H_2,
 \label{eq:benchmark-hamiltonians}
\end{align}
where $H_4$ and $H_2$ contain the random quartic and quadratic interactions of the complex SYK model. Both reservoirs evolve as $U=e^{-i\Delta t H}$ and use complete retained-state readout. We tune $\Delta t$ together with the transverse field $h_z$ for the Ising reservoir, and with the mixing parameter $\alpha$ for the SYK reservoir. Each physical model therefore has two adjustable controls, matching the two parameters $R$ and $\eta$ of the recursive kernel. Model parameters and readout regularization are selected by validation using the same search budget~\cite{SM}.

\begin{figure}[tb]
 \centering
 \includegraphics[width=\columnwidth]{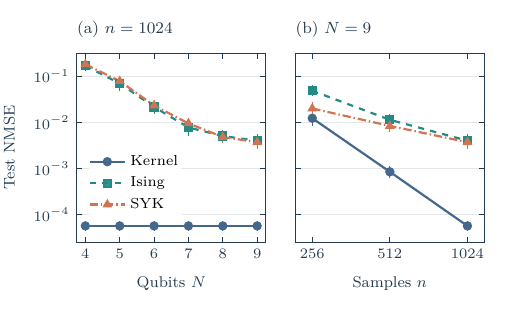}
 \caption{NARMA10 prediction with $x_t\sim\mathrm U[0,0.2]$ and $u_t=5x_t$. (a) Test NMSE versus total qubits at $n=1024$; the recursive-kernel reference is dimension independent. (b) Nine-qubit learning curves ($D=256$) with nested training sets. Points and bars show means and pointwise 95\% bootstrap intervals over eight input/device pairs, conditional on the fixed candidate bank. Input streams are shared across models and sizes.}
 \label{fig:performance}
\end{figure}

With 1024 training examples, increasing the reservoir size from four to nine total qubits steadily improves both Hamiltonian models, while the dimension-independent recursive kernel maintains a lower mean test NMSE throughout this range [Fig.~\ref{fig:performance}(a)]. Fixing the physical reservoirs at nine qubits then allows us to examine the effect of increasing the training data, with parameters reselected at each sample size. The recursive kernel achieves lower mean error at all three training budgets, and its relative error advantage grows as more data become available [Fig.~\ref{fig:performance}(b)]. On NARMA10, the recursive kernel achieves lower mean prediction error than the tested Hamiltonian reservoirs, while its evaluation cost remains independent of reservoir size.

\lettersection{Discussion}The analytical description of reservoir learning extends beyond the Haar setting studied here. We have also identified tractable recursive descriptions in locally driven models of many-body localization and in the long-time limit of large-dimensional Gaussian-unitary-ensemble Hamiltonians. These examples suggest that simple learning kernels can emerge from distinct physical mechanisms, motivating the search for analytical kernels in a wider range of quantum reservoirs. The connection between kernel coefficients and OTOCs offers a further route in this direction. In particular, large-$N$ methods developed for the SYK model~\cite{MaldacenaStanford2016} could be adapted to derive temporal kernels for interacting Hamiltonian reservoirs and reveal how many-body correlations shape their learning properties. More broadly, our results demonstrate that the complexity of quantum dynamics need not carry over to the associated learning model. Establishing quantum advantage in QRC therefore requires a careful assessment of the learning task and its efficient classical alternatives.

\addtocontents{toc}{\protect\begingroup\protect\OmitLetterContents}
\enlargethispage{\baselineskip}

%
\addtocontents{toc}{\protect\endgroup}

\clearpage
\onecolumngrid
\StartSupplementReferences

\setcounter{page}{1}
\renewcommand{\thepage}{S\arabic{page}}
\setcounter{equation}{0}
\setcounter{figure}{0}
\setcounter{table}{0}
\setcounter{section}{0}
\setcounter{secnumdepth}{2}
\numberwithin{equation}{section}
\renewcommand{\thefigure}{S\arabic{figure}}
\renewcommand{\thetable}{S\arabic{table}}
\renewcommand{\thesection}{S\arabic{section}}
\renewcommand{\thesubsection}{\thesection.\Alph{subsection}}
\renewcommand{\theHequation}{supp.\arabic{section}.\arabic{equation}}
\renewcommand{\theHfigure}{supp.\arabic{figure}}
\renewcommand{\theHtable}{supp.\arabic{table}}
\renewcommand{\theHsection}{supp.\arabic{section}}
\renewcommand{\theHsubsection}{supp.\arabic{section}.\arabic{subsection}}
\makeatletter
\renewcommand{\p@subsection}{}
\makeatother
\renewcommand{\citenumfont}[1]{S#1}
\renewcommand{\bibnumfmt}[1]{[S#1]}
\renewcommand{\refname}{Supplemental references}
\begin{center}
{\large\bfseries Supplemental Material}\\[5pt]
{\bfseries Emergent Recursive Kernels in Quantum Reservoir Computing}
\end{center}

We derive the local kernels, establish the recursive Haar limit, and
relate kernel coefficients to operator correlations. We then analyze
finite-dimensional learning effects, computational costs, and the
numerical tests of kernel control and prediction.
Equations, figures, and tables carrying an ``S'' belong to this
Supplemental Material; those without that prefix refer to the
accompanying Letter. Bibliographic citations are numbered separately:
$[1],[2],\ldots$ in the Letter and $[\mathrm S1],[\mathrm S2],\ldots$
here, with the supplemental references collected at the end.

Throughout, $D$ is the retained Hilbert-space dimension, $d$ is the
active retained dimension of a local writer, $L$ is an input-history
length, and $n$ is a training-set size. When the full physical device
contains $N$ qubits, $D=2^{N-1}$ because one qubit is reset at each step.

\setcounter{tocdepth}{2}
\begingroup
\hypersetup{linkcolor=black}
\tableofcontents
\endgroup
\clearpage

\section{Local kernels for injection and memory propagation}
\label{hx:section}

We derive the seven-coefficient form of Eq.~\eqref{eq:seven} from an
arbitrary unitary writer $W_D$ on the input qubit $A$ and the retained
system $R$ of dimension $D$. The two input coordinates determine the
form of the local kernels, while the writer determines their
coefficients and joint constraints. All identities in this section
are exact at finite $D$; Haar randomness enters only in the subsequent
reduction of the temporal dynamics.

\subsection{Input injection}

We use the encoding and input coordinates of the Letter,
\begin{equation}
 \sigma(u)=|\chi(u)\rangle\langle\chi(u)|
 =\frac{I_A+h(u)X-z(u)Z}{2},\qquad
 \boldsymbol s(u)=\begin{pmatrix}h(u)\\z(u)\end{pmatrix},\qquad
 z(u)=2u-1,\quad h(u)=\sqrt{1-z(u)^2}.
 \label{hx:protocol}
\end{equation}
The sign of $Z$ follows from
$|\chi(u)\rangle=\sqrt{1-u}|0\rangle+\sqrt u|1\rangle$.
As in the Letter, $\Delta_u=2\sigma(u)-I_A$ denotes the input operator
\emph{before} the joint evolution. With $P_h=X$ and $P_z=-Z$, it is
$\Delta_u=h(u)P_h+z(u)P_z$.

The injected response is $J_u=\Psi_u(I_D)-I_D$, with $\Psi_u$ defined
in Eq.~\eqref{eq:writer-channel}. Unitarity gives
$\Tr_A[W_D(I_A\otimes I_D)W_D^\dagger]=2I_D$, so the constant part
of the encoded input contributes precisely the identity subtracted
in $J_u$. Only the two input-dependent parts remain:
\begin{equation}
 E_i=\frac12\Tr_A[W_D(P_i\otimes I_D)W_D^\dagger],\quad i=h,z,
 \qquad J_u=h(u)E_h+z(u)E_z.
 \label{hx:source-responses}
\end{equation}
Both $E_h$ and $E_z$ are Hermitian and traceless. Substitution into
the source overlap therefore yields
\begin{equation}
 b_D(u,v)=\frac{\Tr(J_uJ_v)}D
 =\boldsymbol s(u)^{\mathsf T}B\boldsymbol s(v),\qquad
 B_{hh}=\frac{\Tr(E_h^2)}D,\quad
 B_{hz}=B_{zh}=\frac{\Tr(E_hE_z)}D,\quad
 B_{zz}=\frac{\Tr(E_z^2)}D.
 \label{hx:plane}
\end{equation}
These are the three source coefficients directly in terms of the
writer. Their Gram form makes $B$ real symmetric and positive
semidefinite.

\subsection{Memory propagation}

To obtain the propagation coefficients, resolve the same writer in
the input-qubit basis:
\begin{equation}
 W_D=\begin{pmatrix}W_{00}&W_{01}\\W_{10}&W_{11}\end{pmatrix},
 \qquad
 K_\mu(u)=\sqrt{1-u}\,W_{\mu0}+\sqrt u\,W_{\mu1}.
 \label{hx:writer-blocks}
\end{equation}
Each block is a $D\times D$ matrix; its column labels the incoming
qubit state and its row labels the outgoing state. The traces of
block products define two ordinary $2\times2$ matrices,
\begin{equation}
 M_{\mu\nu}=\frac{\Tr(W_{\mu0}W_{\nu0}^\dagger)}D,
 \qquad
 N_{\mu\nu}=\frac{\Tr(W_{\mu0}W_{\nu1}^\dagger)}D,
 \qquad \mu,\nu=0,1.
 \label{hx:block-traces}
\end{equation}
Here $M$ is the outgoing-qubit state for an input $|0\rangle$ and a
maximally mixed retained system; $N$ is the response to the input
coherence $|0\rangle\langle1|$ and need not be Hermitian.
The column and row identities of $W_D^\dagger W_D=W_DW_D^\dagger=I$
give, respectively,
\begin{equation}
 \Tr M=1,\qquad \Tr N=0,\qquad
 \frac{\Tr(W_{\mu1}W_{\nu1}^\dagger)}D
   =(I_2-M)_{\mu\nu}.
 \label{hx:block-identities}
\end{equation}
Thus the responses to all four input matrix elements are
$M,N,N^\dagger,I_2-M$.

These matrices also collect the Kraus overlaps appearing in the
propagation kernel. For two input values define the $2\times2$ matrix
$(\Omega_{uv})_{\mu\nu}=\Tr[K_\mu(u)K_\nu(v)^\dagger]/D$.
It is the outgoing-qubit response to $|\chi(u)\rangle\langle\chi(v)|$
with the retained state $I_D/D$. The original Kraus formula becomes
\begin{equation}
 a_D(u,v)=\frac1{D^2}\sum_{\mu,\nu=0}^1
   |\Tr[K_\mu(u)^\dagger K_\nu(v)]|^2
 =\Tr(\Omega_{uv}\Omega_{uv}^\dagger).
 \label{hx:local}
\end{equation}
Expanding the input amplitudes gives
\begin{equation}
 \Omega_{uv}
 =\sqrt{(1-u)(1-v)}\,M+\sqrt{uv}\,(I_2-M)
  +\sqrt{(1-u)v}\,N+\sqrt{u(1-v)}\,N^\dagger.
 \label{hx:cross-marginal}
\end{equation}
On inserting this expression into Eq.~\eqref{hx:local}, the identities
in Eq.~\eqref{hx:block-identities} cancel the terms proportional to
$h(u)+h(v)$ and $z(u)+z(v)$. Using
$u=(1+z(u))/2$ and $\sqrt{u(1-u)}=h(u)/2$ then yields
\begin{equation}
 a_D(u,v)=a_0+\boldsymbol s(u)^{\mathsf T}C\boldsymbol s(v),
 \label{hx:normal}
\end{equation}
with the four explicit coefficients
\begin{equation}
 \boxed{\begin{aligned}
 a_0&=\frac12\big[\Tr(M^2)+\Tr(NN^\dagger)\big],\\
 C_{hh}&=\frac12\big[1-\Tr(M^2)+\operatorname{Re}\Tr(N^2)\big],\\
 C_{hz}=C_{zh}&=-\operatorname{Re}\Tr(MN),\\
 C_{zz}&=\frac12\big[\Tr(M^2)-\Tr(NN^\dagger)\big].
 \end{aligned}}
 \label{hx:explicit-propagation}
\end{equation}
All traces in Eq.~\eqref{hx:explicit-propagation} are ordinary
$2\times2$ matrix traces. Together with Eq.~\eqref{hx:plane}, these
formulas give all seven real coefficients without additional input
coordinates or a symmetry assumption on $W_D$.

\subsection{Memory--injection tradeoff}

The common unitary origin of these coefficients also restricts their
allowed values. First, the propagation kernel is positive semidefinite,
as can be seen using the reservoir matrix units $e_{rs}=|r\rangle\langle s|$:
\[
 a_D(u,v)=\frac1{D^2}\sum_{r,s=1}^D
 \Tr\big[\Psi_u(e_{rs})^\dagger\Psi_v(e_{rs})\big].
\]
This is a Gram sum of ordinary matrices. Since the functions $1,h,z$
are linearly independent, Eq.~\eqref{hx:normal} implies $a_0\ge0$
and $C\succeq0$, in addition to $B\succeq0$ established above.

To derive the joint constraint on injection and propagation, set
$v=u$ in Eq.~\eqref{hx:cross-marginal}. The matrix $\Omega_{uu}$ is
the outgoing-qubit density matrix
\begin{equation}
 \omega_u=\frac{I_2}{2}
     +\frac{h(u)}2(N+N^\dagger)
     +\frac{z(u)}2(I_2-2M),
 \qquad a_D(u,u)=\Tr(\omega_u^2).
 \label{hx:qubit-marginal}
\end{equation}
The equality follows directly from Eq.~\eqref{hx:local}; it concerns
the writer evaluated with the retained state $I_D/D$.
Now distinguish the evolved input operator from $\Delta_u$ itself:
\begin{equation}
 \widetilde\Delta_u=W_D(\Delta_u\otimes I_D)W_D^\dagger
 =I_A\otimes J_u+(2\omega_u-I_A)\otimes I_D+Q_u.
 \label{hx:local-decomposition}
\end{equation}
The first two terms are its retained-system and input-qubit local
parts. The remainder $Q_u$ has both partial traces zero, so the three
terms have vanishing pairwise trace products. Because the input is
pure, $\Delta_u^2=I_A$ and $\widetilde\Delta_u^2=I_{2D}$.
Squaring Eq.~\eqref{hx:local-decomposition} and taking the normalized
ordinary trace therefore gives
\begin{equation}
 1=b_D(u,u)+\big[2a_D(u,u)-1\big]
       +\frac{\Tr(Q_u^2)}{2D}.
 \label{hx:joint-budget}
\end{equation}
Here $\Tr[(2\omega_u-I_A)^2]/2=2a_D(u,u)-1$.
The last term is nonnegative, proving the injection--propagation
bound. Together with $1/2\le\Tr(\omega_u^2)\le1$, the Gram
Cauchy--Schwarz inequalities and the nonnegative Kraus formula yield
\begin{equation}
 \boxed{b_D(u,u)+2a_D(u,u)\le2},\qquad
 0\le a_D(u,v)\le1,\qquad |b_D(u,v)|\le1.
 \label{hx:bounds}
\end{equation}
Equivalently, the joint diagonal bound constrains the seven coefficients by
\begin{equation}
 B+2C\preceq2(1-a_0)I_2.
 \label{hx:coefficient-budget}
\end{equation}
Indeed, $\boldsymbol s(u)$ covers a half unit circle, and the quadratic
form is unchanged under $\boldsymbol s\mapsto-\boldsymbol s$.
These are necessary physical constraints, not a characterization of
all coefficient choices realizable by a unitary writer.
Replacing $W_D$ by the full update $(I_A\otimes V_D)W_D$ only conjugates
the retained-system factors by $V_D$; it leaves $\omega_u$ and every
trace in Eq.~\eqref{hx:joint-budget} unchanged.

\section{Orthogonal memory and the recursive Haar kernel}
\label{hx:haar-section}

We first determine the leading kernel: asymptotic freeness makes
the current injection orthogonal to the past and reduces memory
propagation to multiplication by a scalar. We then quantify the
finite-dimensional error. The only dimension count needed is how
Haar index matching splits or joins trace loops; it controls both
the discarded terms and the fluctuations of the full kernel.

Throughout, $W_D$ is fixed before drawing
$V_D\sim\mathrm{Haar}[U(D)]$. The same device is used at every step,
and both prescribed input histories start from $\rho_0=I_D/D$.
History lengths are fixed as $D$ grows; all error estimates below
allow constants that depend on those lengths.

\subsection{Exact kernel decomposition}

For the two histories, write $M_t=D\rho_t(\boldsymbol u)-I_D$ and
$M'_t=D\rho_t(\boldsymbol v)-I_D$. Their exact update and kernel are
\begin{equation}
 M_t=V[J_{u_t}+\Psi_{u_t}(M_{t-1})]V^\dagger,\qquad
 k_t=\frac{\Tr(M_tM'_t)}D,\qquad M_0=M'_0=0.
 \label{hx:exact-update}
\end{equation}
Here $V=V_D$, $k_t=k_{D,t}(\boldsymbol u,\boldsymbol v)$, and
$a_t=a_D(u_t,v_t)$, $b_t=b_D(u_t,v_t)$ are the local coefficients
of Sec.~\ref{hx:section}. Cancelling the common final conjugation
and expanding the overlap gives
\begin{align}
 k_t&=b_t+a_tk_{t-1}+c_t+c'_t+r_t,\nonumber\\
 c_t&=\frac{\Tr[J_{u_t}\Psi_{v_t}(M'_{t-1})]}D,\qquad
 c'_t=\frac{\Tr[\Psi_{u_t}(M_{t-1})J_{v_t}]}D,\nonumber\\
 r_t&=\frac{\Tr[\Psi_{u_t}(M_{t-1})\Psi_{v_t}(M'_{t-1})]}D
                  -a_tk_{t-1}.
 \label{hx:exact-residuals}
\end{align}
The first two remainders are the current--past overlaps in
Eq.~\eqref{eq:source-past}; the third is the propagation remainder
in Eq.~\eqref{eq:memory-propagation}. Its subtraction uses the
exact preceding kernel. All three vanish at the first step.

\subsection{Asymptotic freeness and the memory recurrence}

The relevant form of asymptotic freeness can be stated directly
in terms of traces. A writer factor is any fixed polynomial in the
Kraus operators and their adjoints. Organize these factors by the
number of Haar conjugations they have undergone:
\[
 A,\qquad VAV^\dagger,\qquad V^2A(V^\dagger)^2,\qquad\ldots.
\]
At leading order, these families obey the free-moment rule: an
alternating product of centered factors has zero normalized trace
whenever adjacent factors carry different conjugation counts.
Centering means subtracting $\Tr(X)I_D/D$ from $X$; a factor may
itself contain products and sums within one family. Thus, for
centered $X_i$ from one family and centered $Y_i$ from another,
\begin{equation}
 \frac{\Tr(X_1Y_1)}D\simeq0,\qquad
 \frac{\Tr(X_1Y_1X_2Y_2)}D\simeq0.
 \label{hx:free-rule}
\end{equation}
The first relation removes the source--memory overlap. The second
removes the centered part of memory propagation, leaving the
scalar factor derived below. Longer alternating products obey
the same rule. The symbol $\simeq$ denotes equality of the
leading Haar trace terms;
finite-dimensional errors are estimated below. Asymptotic freeness
means that these trace relations become exact in the large-dimension
limit at fixed product length. This is stronger
than pairwise orthogonality, since it also controls mixed products.
It follows from asymptotic freeness of the Haar unitary and the
fixed writer~\cite[Theorem 1.4]{SM-CollinsMale2014}: in the free limit,
expanding factors with different conjugation counts separates
centered writer factors by nonzero powers of the Haar unitary.
The same rule makes the current writer jointly free from all
positive-count families and their combinations at leading order.

The update preserves precisely this structure. At the first step,
$M_1=VJ_{u_1}V^\dagger$ has conjugation count one. At the next step,
\begin{equation}
 M_2=VJ_{u_2}V^\dagger+
 \sum_\mu
 [VK_\mu(u_2)V^\dagger]\,
 [V^2J_{u_1}(V^\dagger)^2]\,
 [VK_\mu(u_2)^\dagger V^\dagger].
 \label{hx:two-step-counts}
\end{equation}
The propagated term therefore contains counts one and two,
although both conjugations use the same $V$. If the entering
memory is built from factors with counts $1,\ldots,t-1$, the current
Kraus operators and injection add count-zero factors. The outer
$V(\cdot)V^\dagger$ then increases every count by one, leaving
$M_t$ built from counts $1,\ldots,t$. Trace preservation keeps
every memory traceless. This induction treats any fixed history
length without expanding its individual paths. In particular, the
two entering memories may be correlated with each other, while
their joint collection remains free from the current writer at
leading order. There is no independent Haar draw at a new time.

Write $M=M_{t-1}$, $M'=M'_{t-1}$ and $u=u_t$, $v=v_t$ for one
update. Cycling the current Kraus operators next to the injection
gives
\begin{equation}
 \frac{\Tr[J_u\Psi_v(M')]}D
 =\sum_\nu\frac{\Tr[K_\nu(v)^\dagger J_uK_\nu(v)M']}D
 \simeq0.
 \label{hx:source-reduction}
\end{equation}
Each summand pairs a current writer factor with a traceless past
memory. The scalar part of the writer factor vanishes against
$\Tr M'=0$, and its centered part vanishes by freeness. The other
current--past overlap is treated identically.

For the overlap of two propagated memories, consider one Kraus
pair and separate the current factor into scalar and traceless parts,
\[
 C=K_\mu(u)^\dagger K_\nu(v)=cI_D+C_0,\qquad
 c=\frac{\Tr C}{D},\qquad \Tr C_0=0.
\]
Writing $k=\Tr(MM')/D$ only within this calculation, the exact expansion is
\begin{align}
 \frac{\Tr(MCM'C^\dagger)}D
 ={}&|c|^2k
 +c^*\frac{\Tr[(M'M-kI_D)C_0]}D
 +c\frac{\Tr[(MM'-kI_D)C_0^\dagger]}D
 \nonumber\\
 &+\frac{\Tr(MC_0M'C_0^\dagger)}D.
 \label{hx:memory-split}
\end{align}
The two terms linear in $C_0$ pair a centered past factor with a
centered current factor: both $M'M-kI_D$ and $MM'-kI_D$ are
traceless. Their leading traces vanish by the two-factor rule.
The last term alternates four centered past and current factors,
and vanishes by the four-factor rule in Eq.~\eqref{hx:free-rule}.
This uses freeness of the current writer from the joint past
collection; the two memories themselves need not be free or
independent. The scalar--scalar term contains their exact
overlap $k$, including all internal historical correlations;
none of these has been discarded or separately approximated.
Consequently,
\begin{equation}
 \frac{\Tr(MCM'C^\dagger)}D
 \simeq\left|\frac{\Tr C}{D}\right|^2\frac{\Tr(MM')}D.
 \label{hx:fourfactor}
\end{equation}
Summing over Kraus pairs gives
\begin{align}
 \frac{\Tr[\Psi_u(M)\Psi_v(M')]}D
 &=\sum_{\mu,\nu}\frac{\Tr[
 M K_\mu(u)^\dagger K_\nu(v)M'K_\nu(v)^\dagger K_\mu(u)]}D
 \nonumber\\
 &\simeq
 \sum_{\mu,\nu}\left|\frac{\Tr[K_\mu(u)^\dagger K_\nu(v)]}{D}\right|^2
                 \frac{\Tr(MM')}D
 =a_D(u,v)\frac{\Tr(MM')}D.
 \label{hx:factorization}
\end{align}
Thus the original overlap between the two histories is retained;
only the effect of the current channel reduces to a scalar.
Both reductions apply at every step of the induction, giving
\begin{align}
 \kappa_{D,0}&=0,\qquad
 \kappa_{D,t}=b_D(u_t,v_t)+a_D(u_t,v_t)\kappa_{D,t-1},\nonumber\\
 \kappa_{D,L}(\boldsymbol u,\boldsymbol v)
 &=\sum_{j=1}^{L}b_D(u_j,v_j)\prod_{s=j+1}^{L}a_D(u_s,v_s).
 \label{hx:proxy}
\end{align}
Memories injected at different times have zero leading overlap;
equal-age memories start with overlap $b_D$ and acquire a factor
$a_D$ at each later update, as illustrated in Fig.~\ref{fig:protocol}.

This completes the leading-order derivation. The history sum is
also a positive-semidefinite kernel at every finite $D$, since
each summand is a product of the local positive-semidefinite
kernels from Sec.~\ref{hx:section}. The symbol $\simeq$ above
identifies the leading structure; it does not yet bound the
finite-dimensional error or its fluctuations for one device.
We next recover these vanishing contributions by explicit Haar
index matching, and extend the argument to second moments. This
both explains the mechanism behind the free-moment reductions
used here and quantifies their finite-dimensional accuracy.

\subsection{Finite-dimensional errors and self-averaging}

We now explain the Haar-index mechanism behind the vanishing
free-moment contributions above, and obtain an additional result:
control of their second moments at finite dimension. The organizing
quantity is the power of $D$ in each Haar contraction. It is fixed
by the following six counts, taken over the entire product being
averaged:
\begin{center}
\begin{tabular}{cll}
\toprule
Count & Meaning & What is counted\\
\midrule
$N_V$ & unitary occurrences & occurrences of $V$ (and equally of $V^\dagger$)\\
$N_{\mathrm i}$ & initial traces & original normalized traces\\
$N_{\mathrm f}$ & final traces & fixed-matrix traces after one Haar contraction\\
$N_{\mathrm s}$ & swaps & minimum swaps implementing that matching\\
$N_{\mathrm j}$ & joins & joins when those swaps act on the original traces\\
$N_1$ & single-factor traces & final traces containing exactly one fixed factor\\
\bottomrule
\end{tabular}
\end{center}
For a mean, $N_{\mathrm i}=1$; for $\E_V|F|^2$,
$N_{\mathrm i}=2$. In both cases all occurrences of the same $V$
are integrated jointly. The proof proceeds through two lemmas:
the first expresses the dimension power in terms of these counts;
the second uses traceless boundary factors to exclude the
order-one contractions. We then apply them to every term of a
finite history.

\begin{smlemma}[Counting relations for Haar contractions]
\label{hx:counting-lemma}
\upshape
Consider a product of $N_{\mathrm i}$ normalized traces, each
containing at least one unitary factor, with $N_V$ occurrences
each of $V$ and $V^\dagger$ in total. Put one fixed
matrix after every unitary occurrence, inserting identities if
needed. Assume fixed word length and uniformly bounded operator
norms for these $2N_V$ fixed factors. Each Haar contraction obeys
\begin{equation}
 N_{\mathrm s}\ge N_V,\qquad
 N_{\mathrm f}=N_{\mathrm i}+N_{\mathrm s}-2N_{\mathrm j},\qquad
 2N_{\mathrm f}-N_1\le2N_V.
 \label{hx:counting-relations}
\end{equation}
Its contribution is bounded by
\begin{equation}
 \underbrace{D^{-N_{\mathrm i}}}_{\text{normalization}}\,
 \underbrace{O(D^{N_{\mathrm f}})}_{\text{final traces}}\,
 \underbrace{O(D^{-N_{\mathrm s}})}_{\text{Haar weight}}
 =O(D^{-2N_{\mathrm j}}).
 \label{hx:loop-count}
\end{equation}
In particular, a contraction with no joins requires
\begin{equation}
 N_{\mathrm j}=0\quad\Longrightarrow\quad
 N_1\ge2N_{\mathrm i}+2(N_{\mathrm s}-N_V)
       \ge2N_{\mathrm i}.
 \label{hx:zero-join-singletons}
\end{equation}
\end{smlemma}

\begin{smproof}
Haar averaging identifies matrix indices. For one pair,
\begin{equation}
 \E_V[V_{ij}\overline{V_{kl}}]=\frac{\delta_{ik}\delta_{jl}}D.
 \label{hx:haar-one-pair}
\end{equation}
Independent row and column phase invariance forces $i=k$ and
$j=l$; permutation invariance makes the remaining expectations
equal, and $\sum_j|V_{ij}|^2=1$ fixes their value. For more pairs,
row and column identifications are chosen separately:
\begin{equation}
 \E_V\!\left[\prod_{p=1}^{N_V}V_{i_pj_p}
                    \overline{V_{i'_pj'_p}}\right]
 =\sum_{\pi_{\rm row},\pi_{\rm col}\in S_{N_V}}
   \left(\prod_{p=1}^{N_V}
   \delta_{i_p,i'_{\pi_{\rm row}(p)}}
   \delta_{j_p,j'_{\pi_{\rm col}(p)}}\right)
   \operatorname{Wg}_D(\pi_{\rm col}^{-1}\pi_{\rm row}).
 \label{hx:haar-integral}
\end{equation}
Here $S_{N_V}$ is the set of rearrangements of $N_V$ labels.
The Weingarten coefficient assigns the weight to each matching;
for fixed $N_V$ it satisfies
\begin{equation}
 \operatorname{Wg}_D(\pi_{\rm col}^{-1}\pi_{\rm row})
 =O\!\left(D^{-N_V-|\pi_{\rm col}^{-1}\pi_{\rm row}|}\right),
 \label{hx:weingarten-order}
\end{equation}
where $|\cdot|$ is the minimum number of swaps of two labels.
These are the standard Haar integration
identities~\cite[Corollary 2.4 and Proposition 2.6]{SM-CollinsSniady2006}.
They do not replace higher Haar moments by products of
Eq.~\eqref{hx:haar-one-pair}.

To translate these index equalities into traces, label the $2N_V$
unitary positions. Write $U_s=V$ at a positive position and
$U_s=V^\dagger$ at a negative position, followed by the fixed
matrix $X_s$. Let $\pi_{\rm in}(s)$ be the next unitary position
in the same original trace. Thus $\pi_{\rm in}$ consists of
$N_{\mathrm i}$ cycles. A cycle $(1\,3\,2)$ means
$1\to3\to2\to1$; maps in a product act from right to left.
Before averaging, the summed matrix indices are
\begin{equation}
 \prod_{s=1}^{2N_V}(U_s)_{a_sb_s}
                       (X_s)_{b_s,a_{\pi_{\rm in}(s)}}.
 \label{hx:uncontracted-indices}
\end{equation}
A negative position contributes
$(V^\dagger)_{a_nb_n}=\overline{V_{b_na_n}}$.
Consequently the row and column deltas together identify
$a_s=b_{\pi_{\rm match}(s)}$, where
\begin{equation}
 \pi_{\rm match}(p)=\pi_{\rm row}(p)\quad\text{at positive positions},
 \qquad
 \pi_{\rm match}(n)=\pi_{\rm col}^{-1}(n)
                         \quad\text{at negative positions}.
 \label{hx:matching-map}
\end{equation}
Here the row and column maps are viewed as bijections from the
positive positions to the negative positions. The matching map
always changes the type of position. The remaining index sum is
\begin{equation}
 \prod_s(X_s)_{b_s,b_{\pi_{\rm out}(s)}},\qquad
 \pi_{\rm out}=\pi_{\rm match}\pi_{\rm in}.
 \label{hx:contracted-indices}
\end{equation}
Each cycle of $\pi_{\rm out}$ now gives an ordinary trace of the
fixed matrices along that cycle. Its cycle count is exactly
$N_{\mathrm f}$; each such trace is at most $D$ times the product
of its factors' operator norms.

We next relate the Haar weight to swaps. For any permutation,
\begin{equation}
 \text{minimum swaps}
 =\text{number of positions}-\text{number of cycles}.
 \label{hx:transposition-length}
\end{equation}
The starting point for this identity is the \emph{identity
permutation}, with every label in a separate cycle. A swap changes
the cycle count by one, giving the lower bound. Conversely, a
cycle of length $k$ can be made with $k-1$ swaps, for example
$(1\,2\,3\,4)=(1\,4)(1\,3)(1\,2)$. Constructing the disjoint
cycles separately attains the bound. Fixed labels count as
one-position cycles and need no swaps.

Apply this identity to $\pi_{\rm match}$. Every cycle alternates
positive and negative positions. Following two steps on its
positive positions gives $\pi_{\rm col}^{-1}\pi_{\rm row}$,
with one cycle for each cycle of $\pi_{\rm match}$. Their common
cycle count is $N_V-|\pi_{\rm col}^{-1}\pi_{\rm row}|$. Hence
\begin{equation}
 N_{\mathrm s}
 =2N_V-\bigl(N_V-|\pi_{\rm col}^{-1}\pi_{\rm row}|\bigr)
 =N_V+|\pi_{\rm col}^{-1}\pi_{\rm row}|\ge N_V.
 \label{hx:matching-length}
\end{equation}
Equation~\eqref{hx:weingarten-order} is therefore $O(D^{-N_{\mathrm s}})$.
The $2N_V$ isolated label cycles used to calculate this minimum
are not the final single-factor traces counted by $N_1$.

Now apply a shortest sequence of these swaps to
\emph{the original trace order $\pi_{\rm in}$}. This second
starting point has $N_{\mathrm i}$ circles, and the endpoint is
$\pi_{\rm out}$. Left multiplication by a swap interchanges two
arrow targets. On the same circle it splits that circle in two;
on different circles it joins them. For example,
\[
 (1\,3)(1\,2\,3\,4)=(1\,2)(3\,4),\qquad
 (2\,4)(1\,2)(3\,4)=(1\,4\,3\,2).
\]
With $N_{\mathrm j}$ joins, the other $N_{\mathrm s}-N_{\mathrm j}$
swaps are splits. Therefore
\begin{equation}
 N_{\mathrm f}=N_{\mathrm i}+N_{\mathrm s}-2N_{\mathrm j},
 \qquad
 N_{\mathrm j}=\frac{N_{\mathrm i}+N_{\mathrm s}-N_{\mathrm f}}2.
 \label{hx:split-join-count}
\end{equation}
In particular, the join count is determined by the endpoints,
independently of the chosen shortest sequence. Intermediate circles
are a counting device, not successive independent Haar averages.
A nonminimal sequence would not give the Haar exponent above.
Combining normalization, final traces, and Haar weight proves
Eq.~\eqref{hx:loop-count}. A join costs two powers of $D$ because
it decreases the number of final traces instead of increasing it.

Finally, of the $N_{\mathrm f}$ final traces, $N_1$ use one factor
each and every other trace uses at least two. Since there are
$2N_V$ factors in total,
\[
 2N_V\ge N_1+2(N_{\mathrm f}-N_1)
              =2N_{\mathrm f}-N_1.
\]
If $N_{\mathrm j}=0$, substituting
$N_{\mathrm f}=N_{\mathrm i}+N_{\mathrm s}$ gives
Eq.~\eqref{hx:zero-join-singletons}.
\end{smproof}

\begin{smlemma}[Traceless boundaries force a join]
\label{hx:boundary-lemma}
\upshape
In the setting of Lemma~\ref{hx:counting-lemma}, a fixed factor
can form a single-factor trace only if its two
neighboring unitary factors have opposite types, $V$ and
$V^\dagger$. Suppose that, in each original trace, all but at
most one of these boundary factors are traceless. Every nonzero
Haar contraction then has
\begin{equation}
 N_1\le N_{\mathrm i},\qquad N_{\mathrm j}\ge1.
 \label{hx:boundary-count}
\end{equation}
For a single nonempty normalized trace word $F(V)$ with equal
numbers of $V,V^\dagger$ and satisfying this boundary condition,
\begin{equation}
 |\E_V F|=O(D^{-2}),\qquad \E_V|F|^2=O(D^{-2}),
 \label{hx:moment-lemma}
\end{equation}
at fixed length and bounded fixed-factor norms.
\end{smlemma}

\begin{smproof}
By Eq.~\eqref{hx:contracted-indices}, $X_s$ forms its own trace
precisely when
\begin{equation}
 \pi_{\rm match}(\pi_{\rm in}(s))=s.
 \label{hx:singleton-condition}
\end{equation}
The matching map changes type, so $s$ and $\pi_{\rm in}(s)$
must have opposite types. These are the positions immediately
before and after $X_s$. A boundary is thus necessary, but the
particular matching must still close this factor onto itself.
An internal Kraus factor between two $V$ factors, or its adjoint
between two $V^\dagger$ factors, cannot form a trace alone.
It can still occur in a trace with other matrices. The restriction
is positional: a current Kraus product at a boundary can form
its own trace.

Isolating any traceless factor makes the contraction exactly zero.
Thus, under the stated assumption, each original trace supplies
at most one possible nonzero single-factor trace, giving
$N_1\le N_{\mathrm i}$. Lemma~\ref{hx:counting-lemma} instead
requires $N_1\ge2N_{\mathrm i}$ if there are no joins.
These inequalities are incompatible. Every surviving contraction
therefore has at least one join and is $O(D^{-2})$.

For the mean of $F$, apply this argument with $N_{\mathrm i}=1$.
For $\E_V|F|^2$, apply it jointly to the two traces $F$ and
$\overline F$, with $N_{\mathrm i}=2$ and $N_V$ counted over
both traces. Conjugation reverses the word and takes adjoints,
preserving its boundary structure and tracelessness. The same
contradiction is now $N_1\le2$ versus $N_1\ge4$.
Cross-matchings between the two traces are included; no
independence is assumed. At fixed word length there are finitely
many matchings, proving both moment bounds.
\end{smproof}

The boundary condition also explains why the retained kernel can
have an order-one contribution. For example, take the original word
\begin{equation}
 \frac1D\Tr(VA V^\dagger C V A'V^\dagger C^\dagger),
 \qquad (X_1,X_2,X_3,X_4)=(A,C,A',C^\dagger).
 \label{hx:four-slot-word}
\end{equation}
The initial order is $\pi_{\rm in}=(1\,2\,3\,4)$. The allowed
matching $\pi_{\rm match}=(1\,4)(2\,3)$ gives
$\pi_{\rm out}=(1\,3)(2)(4)$ and hence
\begin{equation}
 \sum_{b_1,b_2,b_3,b_4}
 A_{b_1b_3}C_{b_2b_2}A'_{b_3b_1}(C^\dagger)_{b_4b_4}
 =\Tr(AA')\Tr C\Tr C^\dagger.
 \label{hx:four-slot-contraction}
\end{equation}
Here $N_V=N_{\mathrm s}=2$, $N_{\mathrm i}=1$,
$N_{\mathrm f}=3$, $N_{\mathrm j}=0$, and $N_1=2$.
The two single-factor traces can survive because $C,C^\dagger$
need not be traceless. Meanwhile traceless $A,A'$ can share a
nonzero trace $\Tr(AA')$. The leading contribution is
$|\Tr C/D|^2\Tr(AA')/D$. Centering either of the isolated
current factors kills this matching exactly. This is why the
boundary lemma applies to the remainders but not to the retained
scalar contribution.

\begin{smtheorem}[Finite-history kernel error]
\label{hx:finite-history-theorem}
\upshape
Fix the writer and both input histories before drawing a single
$V\sim\mathrm{Haar}[U(D)]$, used at every step, and start from
$M_0=M'_0=0$. For fixed Kraus number and the operator-norm bounds
of Sec.~\ref{hx:section}, the local remainders satisfy
\begin{align}
 |\E_V c_t|+|\E_V c'_t|+|\E_V r_t|&=O_t(D^{-2}),\nonumber\\
 \E_V|c_t|^2+\E_V|c'_t|^2+\E_V|r_t|^2&=O_t(D^{-2}).
 \label{hx:local-errors}
\end{align}
For the complete length-$L$ history and the recursive kernel
$\kappa_{D,L}$ in Eq.~\eqref{hx:proxy},
\begin{align}
 |\E_V k_{D,L}-\kappa_{D,L}|&=O_L(D^{-2}),\nonumber\\
 \left(\E_V|k_{D,L}-\kappa_{D,L}|^2\right)^{1/2}
 &=O_L(D^{-1}),\qquad
 \operatorname{Var}_V(k_{D,L})=O_L(D^{-2}).
 \label{hx:theorem}
\end{align}
Here $O_t(D^{-p})$ means a bound $C_tD^{-p}$ for
$D\ge D_0(t)$, with $C_t,D_0(t)$ independent of $D$ but allowed
to depend on the fixed time $t$. The subscript $L$ has the same
meaning for the complete history. No uniform estimate for times
growing with $D$ is asserted.
\end{smtheorem}

\begin{smproof}
We verify the boundary condition for the actual remainder words.
Iterating Eq.~\eqref{hx:exact-update} expands the entering memory as
\begin{equation}
 M=\sum_\alpha P_\alpha A_\alpha P_\alpha^\dagger,
 \qquad
 P_\alpha=VK_{\alpha,1}V\cdots K_{\alpha,p_\alpha-1}V,
 \qquad \Tr A_\alpha=0.
 \label{hx:history-paths}
\end{equation}
Here $A_\alpha$ is an original injection $J$, and $\alpha$ specifies
its time and subsequent Kraus choices. For $p_\alpha=1$,
$P_\alpha=V$. Different paths generally have different lengths;
all use the same $V$. The injection $A_\alpha$ is traceless,
but the propagated path $P_\alpha A_\alpha P_\alpha^\dagger$
need not be traceless separately. At fixed $t$ there are finitely
many paths of bounded length, independently of $D$.
The bounds $\|K_\mu(u)\|_{\rm op},\|J_u\|_{\rm op}\le1$
also bound the fixed factors and coefficients below.

For a source--memory overlap, center the current factor using
the tracelessness of the \emph{complete} memory, before expanding
that memory into paths:
\begin{equation}
 B=\Psi_v^*(J_u)=\sum_\nu K_\nu(v)^\dagger J_uK_\nu(v),
 \qquad B_0=B-\frac{\Tr B}{D}I_D,
 \qquad c_t=\frac{\Tr(B_0M')}D.
 \label{hx:source-centering}
\end{equation}
Each path then contributes $D^{-1}\Tr(PAP^\dagger B_0)$.
Its only boundaries are $A$ and $B_0$, both traceless. The
other source--memory overlap has the same form.

For memory propagation, let $C=K_\mu(u)^\dagger K_\nu(v)=cI_D+C_0$
as in Eq.~\eqref{hx:memory-split}. The scalar term retains the
\emph{exact} preceding overlap $|c|^2\Tr(MM')/D$.
Since $\Tr C_0=0$, the three remaining trace factors are
\begin{equation}
 \frac{\Tr(MC_0M')}D,\qquad
 \frac{\Tr(MM'C_0^\dagger)}D,\qquad
 \frac{\Tr(MC_0M'C_0^\dagger)}D,
 \label{hx:residual-word-types}
\end{equation}
with coefficients $c^*,c,1$. After expanding both memories,
the complete boundary check is shown below. The last column
bounds $N_1$ for a nonzero contraction of the displayed word.
\begin{center}
\begin{tabular}{lll}
\toprule
Trace word (normalization $D^{-1}$ omitted) & Boundary factors & Allowed $N_1$\\
\midrule
$\Tr(PAP^\dagger B_0)$ & $A,B_0$ & $0$\\
$\Tr(PAP^\dagger C_0QA'Q^\dagger I_D)$ & $A,C_0,A',I_D$ & at most $1$\\
$\Tr(PAP^\dagger I_DQA'Q^\dagger C_0^\dagger)$ & $A,I_D,A',C_0^\dagger$ & at most $1$\\
$\Tr(PAP^\dagger C_0QA'Q^\dagger C_0^\dagger)$ & $A,C_0,A',C_0^\dagger$ & $0$\\
\bottomrule
\end{tabular}
\end{center}
All listed boundary factors except $I_D$ are traceless. Internal
Kraus factors lie within a block of $V$ factors or within a block
of $V^\dagger$ factors, so they cannot supply additional
single-factor traces. We keep identities explicit instead of
canceling neighboring unitaries; an identity inside a block is
not a boundary either. Each row therefore satisfies
Lemma~\ref{hx:boundary-lemma}.

For a source path, $N_V=p_\alpha$; for a pair of memory paths,
$N_V=p_\alpha+p'_\beta$. Forming $|F|^2$ doubles the count.
Although these lengths vary, the contradiction
$N_1\le N_{\mathrm i}<2N_{\mathrm i}\le N_1$ for zero joins
is the same. Thus every individual remainder word has mean
$O_t(D^{-2})$ and second moment $O_t(D^{-2})$.
Means add, and the triangle inequality
$\|\sum_j F_j\|_{L^2}\le\sum_j\|F_j\|_{L^2}$, with
$\|F\|_{L^2}=(\E_V|F|^2)^{1/2}$, controls the finite path and
Kraus sums. This proves Eq.~\eqref{hx:local-errors} without
assuming independence between paths or between the two memories.

Finally set $\varepsilon_t=c_t+c'_t+r_t$ and
$\Delta_t=k_t-\kappa_{D,t}$. The exact recurrence gives
\begin{equation}
 \Delta_t=a_t\Delta_{t-1}+\varepsilon_t,\qquad
 \Delta_0=0,\qquad
 \Delta_L=\sum_{t=1}^{L}\varepsilon_t
                         \prod_{s=t+1}^{L}a_s.
 \label{hx:full-error-sum}
\end{equation}
The coefficients are fixed before drawing $V$, and $0\le a_s\le1$
by Eq.~\eqref{hx:bounds}. Linearity of expectation and the same
$L^2$ triangle inequality prove the bias and root-mean-square
bounds in Eq.~\eqref{hx:theorem}, without independence between
times. Since $\kappa_{D,L}$ is deterministic,
$\operatorname{Var}_V(k_{D,L})\le
\E_V|k_{D,L}-\kappa_{D,L}|^2$, proving the variance bound.
\end{smproof}

As a finite-dimensional check, at the second update
$M_1=VJ_{u_1}V^\dagger$, $M'_1=VJ_{v_1}V^\dagger$, and
$k_1=b_D(u_1,v_1)$ is nonrandom. For $N_V=2$, the Haar weights
are $\operatorname{Wg}_D(e)=1/(D^2-1)$ and
$\operatorname{Wg}_D((1\,2))=-1/[D(D^2-1)]$. They give
\begin{equation}
 \E_V\frac{\Tr(M_1CM'_1C^\dagger)}D
 =\left[|c|^2-
 \frac{\Tr(C_0C_0^\dagger)/D}{D^2-1}\right]k_1,
 \qquad D>1.
 \label{hx:memory-two-step}
\end{equation}
This exhibits the retained coefficient and its $D^{-2}$ correction.
The exactly extracted scalar term need not equal a single
zero-join contraction at finite $D$, since the weight of an
individual contraction also has finite-dimensional corrections.

The theorem further implies, for prescribed histories and
$\varepsilon>0$,
\begin{equation}
 \Pr_V\{|k_{D,L}-\kappa_{D,L}|\ge\varepsilon\}
 \le\frac{C_L}{D^2\varepsilon^2}.
 \label{hx:proxy-tail}
\end{equation}
Under the stated bounds on Kraus number and operator norms,
$C_L$ and the sufficiently-large-$D$ threshold can be chosen
uniformly over the writer and each prescribed history pair.
These are absolute-error estimates, including when the leading
overlap vanishes; they do not control the largest error over
histories selected after observing $V$. A random writer independent
of $V$ is allowed by conditioning on it.

The estimates use local coefficients at the actual dimension.
When these are held fixed, as for $W_D=w\otimes I_m$, the recurrence
gives a dimension-independent Haar limit. A prescribed warm-up
starting from $I_D/D$ with the same device can be included in the
fixed-length histories. Its contribution is retained; increasing
$D$ alone does not erase it or establish an echo-state property
for arbitrary initial states.

\subsection{Factorization freedom and the independence assumption}
\label{hx:independence-gauge}

For any retained-system unitary $Q_D$, the transformation
\begin{equation}
 V_D'=V_DQ_D,\qquad W_D'=(I_A\otimes Q_D^\dagger)W_D
 \label{hx:gauge}
\end{equation}
preserves $U=(I_A\otimes V_D)W_D$. Since
$K_\mu'=Q_D^\dagger K_\mu$ and $J_u'=Q_D^\dagger J_uQ_D$,
it also preserves $a_D,b_D$, the exact kernel, and the recursive proxy.
This raises an apparent paradox: starting from a local evolution
$U=W_{\rm loc}$, any Haar-sampled $V_D$ gives the factorization
$U=(I_A\otimes V_D)\widetilde W_D$ with
$\widetilde W_D=(I_A\otimes V_D^\dagger)W_{\rm loc}$. The theorem
would then seem to apply even when the physical dynamics remains confined
to a fixed local block. The construction, however, chooses the writer
after observing the router and makes them exactly dependent. The theorem
instead fixes the writer before independently sampling the router. Once
$\widetilde W_D$ is fixed, reproducing that prescribed local evolution
requires a uniquely matched router, a zero-measure event under independent
Haar sampling.

The failure under correlated sampling can be exhibited at every
dimension. Let $D=2m$, let $S_D$ swap the input with one retained qubit,
and choose
\begin{equation}
 W_D(V_D)=(I_A\otimes V_D^\dagger)S_D.
 \label{hx:correlated-writer}
\end{equation}
The full update is then always $S_D$. After each input the retained
state is $\sigma(u_t)\otimes I_m/m$, so the two-step diagonal kernel
is $k_{D,2}=1$. Its local coefficients are $b_D(u,u)=1$ and
$a_D(u,u)=1/2$, giving $\kappa_{D,2}=3/2$. The error $1/2$ does not
decrease with dimension.

For an independently sampled router, the probability of a prescribed finite discrepancy instead vanishes according to Eq.~\eqref{hx:proxy-tail}.
Absorbing a previously sampled router into its writer leaves this
devicewise discrepancy unchanged, but produces an induced random-writer
ensemble with identity router. It does not extend the theorem to
arbitrary deterministic writers with identity router or to writers
selected after the Haar draw.

\subsection{Numerical tests of the Haar limit}
\label{smfinite:checks}
\label{smfinite:local-reductions}

For Fig.~\ref{fig:protocol}(c,d), set $Q=Z\otimes I_{D/2}$,
$G=X\otimes I_{D/2}$, and $c_\pm=\sqrt{(1\pm r)/2}$ with $r=0.7$.
We use the writer
\begin{equation}
 W_D(r)=\frac1{\sqrt2}
 \begin{pmatrix}
 c_+I-c_-Q&c_+I+c_-Q\\
 G(c_-I+c_+Q)&G(c_-I-c_+Q)
 \end{pmatrix}.
 \label{hx:pauli-writer}
\end{equation}
Its local kernels follow from Eqs.~\eqref{hx:plane}
and~\eqref{hx:local} and are unchanged across dimensions.

We use 128 common length-$16$ histories with iid inputs
$u_t\sim\mathrm U[0,1]$, each starting from $I_D/D$, at
$D=4,8,16,32,64$. At each dimension, 48 independent Haar routers
are drawn, each fixed throughout all steps and histories.

Panel (c) evaluates the source--memory overlaps and propagation
residual in Eq.~\eqref{hx:exact-residuals} at the final step,
using the exact preceding kernel. Panel (d) evaluates the full-kernel
difference $k_{D,L}-\kappa_{D,L}$. Squared errors are averaged over
all $128^2$ ordered history pairs, including diagonal pairs, and
over routers before taking the square root; the two source--memory
terms are averaged together. Faint points in panel (d) show the
corresponding RMS error over history pairs for each router.
Pointwise 95\% percentile-bootstrap intervals use 2000 resamples of
routers with the histories held fixed. Dotted lines indicate
$D^{-1}$ scaling.

\section{Operator correlations underlying the temporal kernel}
\label{smotoc:section}

A history feature contributes to the kernel through the operator
response it leaves in the reservoir. We show how the overlaps of
these responses become explicit multitime correlations, beginning
with a single-input coefficient that is a Pauli-averaged
out-of-time-ordered correlator (OTOC). Throughout this section,
$U\in U(2D)$ is an arbitrary fixed joint unitary, reused at every
step and for both histories, and the retained system starts from
$I_D/D$. The identities below are exact at finite $D$.

\subsection{Injection kernel and OTOCs}

Consider the coefficient of $z(u)z(v)$ in the one-step kernel.
With the encoding of Eq.~\eqref{hx:protocol}, the corresponding
retained-system response and its squared norm are
\begin{equation}
 E_z^U=-\frac12\Tr_A(UZ_AU^\dagger),\qquad
 B_{zz}(U)=\frac{\Tr_R[(E_z^U)^2]}D.
 \label{smotoc:source-response}
\end{equation}
All Pauli operators act on the input qubit $A$ and, in joint-system
expressions, include $I_D$. The superscript $U$ distinguishes this
response from the writer response in Sec.~\ref{hx:section}.

The partial trace can be eliminated using Pauli completeness.
Writing $\sigma_A^c\in\{I_A,X_A,Y_A,Z_A\}$, for any joint operators
$Q,S$ one has
\begin{align}
 \frac{I_A}{2}\otimes\Tr_A Q
   &=\frac14\sum_{c=0,x,y,z}\sigma_A^c Q\sigma_A^c,\nonumber\\
 \Tr_R[(\Tr_A Q)(\Tr_A S)]
   &=\frac12\sum_{c=0,x,y,z}
           \Tr[Q\sigma_A^c S\sigma_A^c].
 \label{smotoc:twirl}
\end{align}
The second identity follows by multiplying the first by $S$ and
taking the joint trace. This is the same completeness step that
relates subsystem purity to averaged OTOCs~\cite{SM-Fan2017OTOC}.
Applying it to the two copies of $UZ_AU^\dagger$ gives
\begin{equation}
 \boxed{
 B_{zz}(U)=\frac14\sum_{c=0,x,y,z}
 \frac{\Tr[Z_A\sigma_A^c(1)Z_A\sigma_A^c(1)]}{2D},
 \qquad \sigma_A^c(j)=U^{-j}\sigma_A^cU^j.}
 \label{smotoc:local-otoc}
\end{equation}
Here $j$ counts unitary steps. For $U=e^{-iH\Delta t}$, the four operators alternate between times
$0$ and $\Delta t$. Each nontrivial term is an infinite-temperature
four-point OTOC; the identity Pauli supplies the trivial term.
The normalization and physical meaning are transparent in two limits.
For $U=I$, the four terms are $1,-1,-1,1$, so $B_{zz}=0$:
the input leaves no response in the retained system. For a SWAP
between the input and one retained qubit, all four terms equal one
and $B_{zz}=1$, corresponding to complete transfer of the input's
$z$ component.

The same calculation gives the full source kernel. With
$\Delta_u=2\sigma(u)-I_A$ as in the Letter and
$J_u^U=\frac12\Tr_A(U\Delta_uU^\dagger)$,
\begin{equation}
 b^U(u,v)=\frac{\Tr_R(J_u^UJ_v^U)}D
 =\frac14\sum_{c=0,x,y,z}
   \frac{\Tr[\Delta_u\sigma_A^c(1)\Delta_v\sigma_A^c(1)]}{2D}.
 \label{smotoc:source-otoc}
\end{equation}
Identical inputs give an OTOC average; distinct inputs give its
mixed four-point counterpart. For $U=(I_A\otimes V_D)W_D$,
the retained-system rotation cancels from this trace, so
$b^U=b_D$. This establishes Eq.~\eqref{eq:one-step-otoc} without
a Haar average.

\subsection{History coefficients and multitime correlations}

Longer histories follow by resolving each input into its three
coordinates. Set $s_0=1$, $s_h=h$, $s_z=z$ and
$P_0=I$, $P_h=X$, $P_z=-Z$, so that
$\sigma(u)=\frac12\sum_i s_i(u)P_i$.
A label $\alpha=(\alpha_1,\ldots,\alpha_L)\in\{0,h,z\}^L$ specifies
the history function
\begin{equation}
 m_\alpha(\boldsymbol u)=\prod_{t=1}^L s_{\alpha_t}(u_t).
 \label{smotoc:history-functions}
\end{equation}
Time runs from oldest to newest: $(z,0,h)$ labels
$z(u_1)h(u_3)$. A zero label selects the input's identity component;
it does not set that input to zero or omit its evolution step.

The operator response is obtained by selecting the same component
at each update:
\begin{equation}
 H_\varnothing^U=I_D,\qquad
 H_{(\alpha_1,\ldots,\alpha_t)}^U
 =\frac12\Tr_A\!\left[
 U(P_{\alpha_t}\otimes H_{(\alpha_1,\ldots,\alpha_{t-1})}^U)
 U^\dagger\right].
 \label{smotoc:responses}
\end{equation}
Linearity in the input state now yields
\begin{align}
 D\rho_L(\boldsymbol u)-I_D
   &=\sum_{\alpha\ne0^L}m_\alpha(\boldsymbol u)H_\alpha^U,\nonumber\\
 k_{D,L}(\boldsymbol u,\boldsymbol v)
   &=\sum_{\alpha,\beta\ne0^L}
       \mathcal C_{\alpha\beta}^{(L)}(U)
       m_\alpha(\boldsymbol u)m_\beta(\boldsymbol v),\qquad
 \mathcal C_{\alpha\beta}^{(L)}(U)
   =\frac{\Tr_R(H_\alpha^UH_\beta^U)}D.
 \label{smotoc:gram}
\end{align}
Both labels in the kernel sum exclude $0^L$.
Indeed, the all-identity response is $I_D$, whereas every other
response is traceless: taking the trace of one update multiplies
the preceding trace by $\Tr_A(P_i)/2=\delta_{i0}$.
Thus the subtraction removes exactly the constant response.
Since $1,h,z$ are linearly independent, their tensor products define
unique coefficients. The matrix $\mathcal C$ is their real
positive-semidefinite response Gram matrix, distinct from the local
$2\times2$ propagation matrix $C$ in Sec.~\ref{hx:section}.
Its diagonal entries measure response strengths; its off-diagonal
entries measure overlaps between different responses and may have
either sign.

To express these coefficients directly as dynamical correlations,
apply Eq.~\eqref{smotoc:twirl} to every intermediate partial trace
and to the final overlap. The result, for nonzero $\alpha,\beta$, is
\begin{equation}
\begin{aligned}
 \mathcal C_{\alpha\beta}^{(L)}(U)
 ={}&\frac{1}{2D\,4^{2L-1}}
 \sum_{c,\boldsymbol\mu,\boldsymbol\nu}
 \Tr\Bigg[
 U P_{\alpha_L}
 \left(\prod_{j=L-1\downarrow1}\sigma_A^{\mu_j}UP_{\alpha_j}\right)
 \left(\prod_{j=1\uparrow L-1}U^\dagger\sigma_A^{\mu_j}\right)
 U^\dagger\sigma_A^c\\[-2pt]
 &\hspace{25mm}\times
 U P_{\beta_L}
 \left(\prod_{j=L-1\downarrow1}\sigma_A^{\nu_j}UP_{\beta_j}\right)
 \left(\prod_{j=1\uparrow L-1}U^\dagger\sigma_A^{\nu_j}\right)
 U^\dagger\sigma_A^c
 \Bigg].
 \label{smotoc:exact-word}
\end{aligned}
\end{equation}
The arrows specify left-to-right multiplication in the indicated
order; empty products equal the identity. Each of the $L-1$ indices
in $\boldsymbol\mu$ and $\boldsymbol\nu$, and the final index $c$,
is independently summed over $0,x,y,z$. The $P_i$ insertions carry
the signed history labels, while all summed $\sigma_A^c$ are
standard Pauli operators.

There are $L-1$ internal resets in each history and one final
overlap, giving $2L-1$ Pauli sums. More explicitly, the first
identity in Eq.~\eqref{smotoc:twirl} gives
$H_\alpha^U=(2\,4^{L-1})^{-1}\sum_{\boldsymbol\mu}
\Tr_A[\cdots]$, using the first history's ordered string in
Eq.~\eqref{smotoc:exact-word}, ending at $U^\dagger$ before the final
$\sigma_A^c$. Applying the second identity
to the product of these two responses gives the prefactor
$(2D\,4^{2L-1})^{-1}$. This proves the formula using the same
physical $U$ throughout, without a statistical approximation.
An initial string of zero labels can be removed because it leaves
$I_D$ unchanged; zeros after the first nonzero label generally cannot.

Two-step examples make the connection to higher-order, multitime
OTOCs explicit. The coefficient $\mathcal C^{(2)}_{(z,h),(z,h)}$ multiplies
$z(u_1)h(u_2)z(v_1)h(v_2)$ and measures the response to a
two-time product feature. With $X_A(1)=U^\dagger X_AU$,
Eq.~\eqref{smotoc:exact-word} and cyclicity of the trace give
\begin{equation}
\begin{aligned}
 \mathcal C^{(2)}_{(z,h),(z,h)}
 ={}&\frac1{64}\sum_{c,\mu,\nu}\frac1{2D}
 \Tr\big[
 X_A(1)\sigma_A^\mu(1)Z_A\sigma_A^\mu(1)\sigma_A^c(2)\\[-2pt]
 &\hspace{27mm}\times
 X_A(1)\sigma_A^\nu(1)Z_A\sigma_A^\nu(1)\sigma_A^c(2)\big].
 \label{smotoc:two-step-examples}
\end{aligned}
\end{equation}
Replacing both $X_A(1)$ insertions by the identity gives
$\mathcal C^{(2)}_{(z,0),(z,0)}$, the strength of the first input's
response after one further reset and evolution.

The overlap of this older memory with the current injection is
$\mathcal C^{(2)}_{(z,0),(0,z)}$, the coefficient of $z(u_1)z(v_2)$.
Since $H_{(0,z)}^U=E_z^U$, removing its initial identity step gives
\begin{equation}
 \mathcal C^{(2)}_{(z,0),(0,z)}
 =\frac1{16}\sum_{\mu,c}
 \frac{\Tr[
 \sigma_A^\mu(1)Z_A\sigma_A^\mu(1)
 \sigma_A^c(2)Z_A(1)\sigma_A^c(2)]}{2D},
 \qquad Z_A(1)=U^\dagger Z_AU.
 \label{smotoc:cross-age-example}
\end{equation}
All Pauli indices are summed independently over $0,x,y,z$.
The time labels refer to unitary Heisenberg evolution, while the
Pauli insertions and sums retain the intermediate reset.
These Pauli averages include out-of-time-ordered correlations as well
as terms that reduce to ordinary or trivial correlations.

\section{Finite-dimensional effects on learning}
\label{smfc:section}
\label{smgeom:section}

We first show that finite-device readouts belong to the corresponding
Haar feature space. We then compare the minimum readout weights needed
for the same target and use a common error bound to relate their ratio
to sufficient sample budgets.

\subsection{Finite-dimensional readouts in the Haar feature space}

Fix $W_D$ and any router $V\in U(D)$. Every history starts from $I_D/D$
and contains $L$ inputs. The proxy $\kappa_{D,L}$ uses this writer's
actual finite-$D$ local kernels. With $J_u=\Psi_u(I_D)-I_D$,
both local kernels are response Grams:
\begin{align}
 b_D(u,v)&=\frac{\Tr(J_uJ_v)}D
          =\phi_b(u)\phi_b(v)^{\mathsf T},\nonumber\\
 a_D(u,v)&=\frac1{D^2}\sum_{r,s=1}^D
     \Tr[\Psi_u(|r\rangle\langle s|)^\dagger
          \Psi_v(|r\rangle\langle s|)]
          =\phi_a(u)\phi_a(v)^{\mathsf T}.
 \label{smfc:supergram}
\end{align}
Here $r,s$ run over an orthonormal reservoir basis; expanding the Kraus
operators recovers Eq.~\eqref{hx:local}. Choose input-independent
orthonormal bases $O_j$ and $\mathcal A_i$ for the real spans of the
source responses and channels, using these inner products. Then
$J_u=\sum_j\phi_{b,j}(u)O_j$ and
$\Psi_u(M)=\sum_i\phi_{a,i}(u)\mathcal A_i(M)$.
The $\mathcal A_i$ are fixed linear maps, not necessarily quantum
channels themselves: all input dependence enters through the same
scalar functions $\phi_{a,i}(u)$ that define $a_D$.
Minimal response spans handle singular local Grams: a zero Gram
direction has zero response norm. All feature vectors below are rows.

Expanding the scalar recurrence defines its temporal features:
\begin{equation}
 \Phi_t=\phi_b(u_t)\oplus[\phi_a(u_t)\otimes\Phi_{t-1}],\quad
 \Phi_0=0,\qquad
 \kappa_{D,t}(\boldsymbol u,\boldsymbol v)
 =\Phi_t(\boldsymbol u)\Phi_t(\boldsymbol v)^{\mathsf T}.
 \label{smfc:dictionary}
\end{equation}
The same factors occur in the physical update. If
$M_{t-1}=\sum_r\Phi_{t-1,r}R_{t-1,r}$, then
\begin{align}
 M_t&=\sum_j\phi_{b,j}(u_t)VO_jV^\dagger
       +\sum_{i,r}\phi_{a,i}(u_t)\Phi_{t-1,r}
          V\mathcal A_i(R_{t-1,r})V^\dagger,
 \label{smfc:state-induction}\\
 F(\boldsymbol u)&=\Phi_L(\boldsymbol u)T_D^{\rm raw},\qquad
 k_{D,L}(\boldsymbol u,\boldsymbol v)
 =\Phi_L(\boldsymbol u)T_D^{\rm raw}(T_D^{\rm raw})^{\mathsf T}
      \Phi_L(\boldsymbol v)^{\mathsf T}.
 \label{smfc:metric}
\end{align}
Here $M_t=D\rho_t-I_D$. Starting from $M_0=0$ proves the first identity
inductively with fixed traceless Hermitian responses $R_{L,r}$.
Each $\mathcal A_i$ preserves tracelessness because it is a real linear
combination of trace-preserving channels.
In the second, $F$ contains coordinates in a traceless Hermitian basis
$Q_\ell$ satisfying $\Tr(Q_\ell Q_m)/D=\delta_{\ell m}$, and
$(T_D^{\rm raw})_{r\ell}=\Tr(Q_\ell R_{L,r})/D$.
Every linear readout is therefore a combination of Haar features:
$\mathcal H_{k_{D,L}}\subseteq\mathcal H_{\kappa_{D,L}}$.
This is exact for every router, without statistical orthogonality.
The response matrices can overlap, vanish, or be linearly dependent;
these possibilities do not introduce functions outside the Haar dictionary.

Centering on a common input law preserves the transfer identity;
an intercept handles the removed mean. A continuing stream instead
requires its boundary state or a forgetting bound. Likewise, replacing
the actual local kernels by limiting coefficients requires checking
that no necessary feature direction disappears.

\subsection{Relative readout weights}

A target represented by both kernels may nevertheless require very
different readout weights. We quantify the largest such difference
over targets readable by the finite-dimensional reservoir.
For a fixed history distribution $P$, let $\bar k_D,\bar k_H$ denote
the two nonzero kernels centered under $P$ and normalized to
$\mathbb E_P\bar k_j(\boldsymbol u,\boldsymbol u)=1$.
Writing their feature maps as $\bar\phi_j$, define the minimum squared
weights for the same centered target $f$ and their maximal ratio:
\begin{align}
 s_j(f)&=\min_{f=\bar\phi_jw}\|w\|^2
        =\|f\|_{\mathcal H(\bar k_j)}^2,
 \label{smgeom:minimum-norm}\\
 \Gamma_D&=\max_{0\ne f\in\mathcal H(\bar k_D)}
             \frac{s_H(f)}{s_D(f)}.
 \label{smgeom:ratio}
\end{align}
Here $j\in\{D,H\}$, and equality of functions is on
$\operatorname{supp}P$. Minimizing over weights removes redundant
coefficients.

This ratio can be computed directly from kernel matrices~\cite{SM-Huang2021Power}.
Choose a common probe set $\mathcal X\subseteq\operatorname{supp}P$ whose function values uniquely
determine every function in the centered Haar space. Form
$\mathbf K_k,\mathbf K_\kappa$ from the normalized kernels above,
without further empirical centering or rescaling on these probes.
Completeness makes the minimum interpolation weights equal to the
minimum weights for the full target function, so for $\mathbf y=f(\mathcal X)$,
\begin{equation}
 s_D(f)=\mathbf y^{\mathsf T}\mathbf K_k^+\mathbf y,\qquad
 s_H(f)=\mathbf y^{\mathsf T}\mathbf K_\kappa^+\mathbf y,
 \label{smgeom:complete-probe-norms}
\end{equation}
where $+$ denotes the Moore--Penrose pseudoinverse.
To maximize their ratio, write
$\mathbf y=\mathbf K_k^{1/2}\mathbf v$ with
$\mathbf v\in\operatorname{range}\mathbf K_k$. Then
\[
 \mathbf y^{\mathsf T}\mathbf K_k^+\mathbf y
 =\mathbf v^{\mathsf T}
   \mathbf K_k^{1/2}\mathbf K_k^+\mathbf K_k^{1/2}\mathbf v
 =\mathbf v^{\mathsf T}\mathbf v,
\]
because the middle matrix is the orthogonal projector onto that range.
The numerator transforms in the same way, yielding
\begin{align}
 \Gamma_D
 &=\max_{\substack{\mathbf y\ne0\\
                  \mathbf y\in\operatorname{range}\mathbf K_k}}
   \frac{\mathbf y^{\mathsf T}\mathbf K_\kappa^+\mathbf y}
        {\mathbf y^{\mathsf T}\mathbf K_k^+\mathbf y}\nonumber\\
 &=\max_{\substack{\mathbf v\ne0\\
                  \mathbf v\in\operatorname{range}\mathbf K_k}}
   \frac{\mathbf v^{\mathsf T}
     (\mathbf K_k^{1/2}\mathbf K_\kappa^+\mathbf K_k^{1/2})\mathbf v}
        {\mathbf v^{\mathsf T}\mathbf v}\nonumber\\
 &=\lambda_{\max}\!\left(
   \mathbf K_k^{1/2}\mathbf K_\kappa^+\mathbf K_k^{1/2}\right).
 \label{smgeom:complete-probe-ratio}
\end{align}
The last step maximizes a Rayleigh quotient. Its symmetric
positive-semidefinite matrix vanishes on the orthogonal complement
of $\operatorname{range}\mathbf K_k$, so the restricted maximum is
its full largest eigenvalue.
Incomplete probes instead compare interpolation weights and need not
recover $\Gamma_D$. The probes are separate from the learning data.

\subsection{Expected test error and sufficient sample budgets}

To connect readout weights to data requirements, fix a nonzero centered
target $f$ readable by the finite-dimensional reservoir before drawing
independent training pairs with inputs from $P$.
Labels equal $f$ plus conditionally zero-mean noise with a common
finite bound on its conditional variance.
Let $\widehat f_D$ and $\widehat f_H$ be measurable least-squares
empirical minimizers for the normalized finite and Haar kernels,
with squared readout-weight budgets $s_D(f)$ and $s_H(f)$, respectively.
The admissible predictions and $f$ obey a common fixed
amplitude bound on $\operatorname{supp}P$. A standard complexity
argument~\cite[proof of Theorem~8, Theorem~12(4), and
Lemma~22]{SM-BartlettMendelson2002}, together with a second-moment bound
on the noise, gives
\begin{equation}
 \mathbb E_{\rm train}\operatorname{MSE}(\widehat f_D,f)
 \le C\sqrt{\frac{s_D(f)}{n}},\qquad
 \mathbb E_{\rm train}\operatorname{MSE}(\widehat f_H,f)
 \le C\sqrt{\frac{s_H(f)}{n}}.
 \label{smgeom:mse-bound}
\end{equation}
Here MSE is the population error against the noiseless target $f$.
The constant $C$ depends only on the common amplitude and noise bounds;
it is the same for both kernels and independent of $n$ and reservoir size.
Thus, larger squared readout weights require a proportionally larger
sample budget to obtain the same error guarantee.

To see the scaling, empirical optimality against the admissible target
bounds the expected clean error by a sampling fluctuation and a noise
correlation. Symmetrization and contraction control the first;
independence, conditional zero mean, and the variance bound control
the second. Unit mean squared feature length gives the displayed
dependence on weights and sample size. This argument requires no
bounded-noise assumption.

For a common error tolerance $\varepsilon>0$, define the continuous
sufficient sample budgets for the finite reservoir and the Haar kernel,
using the notation of the Letter:
\begin{equation}
\begin{gathered}
 n_D(\varepsilon):=\frac{C^2s_D(f)}{\varepsilon^2},\qquad
 n_H(\varepsilon):=\frac{C^2s_H(f)}{\varepsilon^2},\\[3pt]
 \frac{n_H}{n_D}=\frac{s_H(f)}{s_D(f)}\le\Gamma_D.
\end{gathered}
 \label{smgeom:budget}
\end{equation}
Training sizes of at least $\lceil n_D\rceil$ for the finite reservoir
and $\lceil n_H\rceil$ for the Haar kernel guarantee the stated expected MSE.
The ratio follows from these definitions and the common $C$;
its upper bound is the maximal weight ratio of
Eq.~\eqref{smgeom:complete-probe-ratio}.
These thresholds concern the specified constrained learner with its
weight budget supplied, not the minimum samples needed in practice.
The validation-selected ridge fits are assessed separately in
Sec.~\ref{smrw:section}. Figure~\ref{fig:finite}(a) averages the
devicewise bounds $\Gamma_D$, whereas panel (b) uses measured
attainment budgets. The theoretical budget ratio does not by itself
bound the ratio of actual sample requirements.

\section{Computational cost of reservoir learning}
\label{smrt:section}
\label{smrw:section}

The numerical protocols underlying Figs.~\ref{fig:protocol}--\ref{fig:performance}
are summarized in Table~\ref{sm:protocol-overview}; detailed procedures
are given in the corresponding sections.

\begin{table}[h]
\centering
\small
\caption{Ensembles and data repetitions used in the four main-text
figures. The uncertainty estimators and their conditioning are specified
in the corresponding sections.}
\label{sm:protocol-overview}
\begin{tabular}{@{}p{.11\textwidth}p{.23\textwidth}p{.29\textwidth}p{.29\textwidth}@{}}
\toprule
\raggedright Figure & \raggedright Question & \raggedright Histories and targets & \raggedright Ensembles and repetitions \tabularnewline
\midrule
\raggedright 1 & \raggedright Haar reduction & \raggedright Reset histories, $L=16$; state overlaps, no regression labels & \raggedright 48 routers per $D$, with 128 common histories \tabularnewline[3pt]
\raggedright 2 & \raggedright Learning and total cost & \raggedright Reset histories, $L=8$; device-dependent readouts with training/validation noise variance $0.1$ & \raggedright 12 writers and 4 routers per writer, family, and $D$; 3 data/noise repetitions per device \tabularnewline[3pt]
\raggedright 3 & \raggedright Classical kernel controls & \raggedright Reset histories, $L=8$; orthogonal polynomial targets with training/validation noise variance $0.1$ & \raggedright 8 paired datasets; all specified lag and order targets \tabularnewline[3pt]
\raggedright 4 & \raggedright NARMA10 prediction & \raggedright Continuous streams with washout and gaps; no added label or measurement noise & \raggedright 8 paired streams and device realizations; a fixed candidate bank \tabularnewline
\bottomrule
\end{tabular}
\end{table}

All regression experiments in Figs.~\ref{fig:finite}(b),
\ref{fig:design}, and~\ref{fig:performance} center features and labels
using the training set, normalize the centered training kernel to
unit mean diagonal, and retain these training statistics for validation
and testing, with the training-label mean restored as an unpenalized
intercept.

Figure~\ref{fig:finite}(b) compares the operations needed to reach a
common prediction error. For each device and model, the attained
sample size is measured first; the cost of one fit at that size is
then evaluated. Both main-text curves use the full training kernel.
The following accounting includes classical state simulation and
distinguishes this algorithm from equivalent feature-space regression.

\subsection{Kernel evaluation and regression costs}

For a dense router, propagating one length-$L$ history requires two
complex $D\times D$ products per step, or $O(LD^3)$ operations.
A fixed local writer adds $O(LD^2)$ work. Once the states are known,
each overlap costs $O(D^2)$; the Haar recurrence instead computes a
kernel entry directly in $O(L)$ operations. Both methods then
diagonalize an $n\times n$ training kernel. One decomposition serves
all ridge penalties because regularization only shifts the eigenvalues.

\begin{table}[htb]
\centering
\small
\caption{Leading costs at fixed local writer dimension. Here $n$ is
the model's attained training size, $n_v$ the validation size, and
$m$ the number of candidate ridge penalties.}
\label{smrw:cost-table}
\begin{tabular}{@{}p{.38\textwidth}cc@{}}
\toprule
Contribution & Finite QRC & Haar \\
\midrule
State simulation & $O((n+n_v)LD^3)$ & Not required \\
Training and validation kernels & $O((n+n_v)nD^2)$ & $O((n+n_v)nL)$ \\
One eigendecomposition & $O(n^3)$ & $O(n^3)$ \\
Fits and validation & $O(m[n^2+n_vn])$ & $O(m[n^2+n_vn])$ \\
\bottomrule
\end{tabular}
\end{table}

\begin{figure*}[t]
\centering
\includegraphics[width=\textwidth]{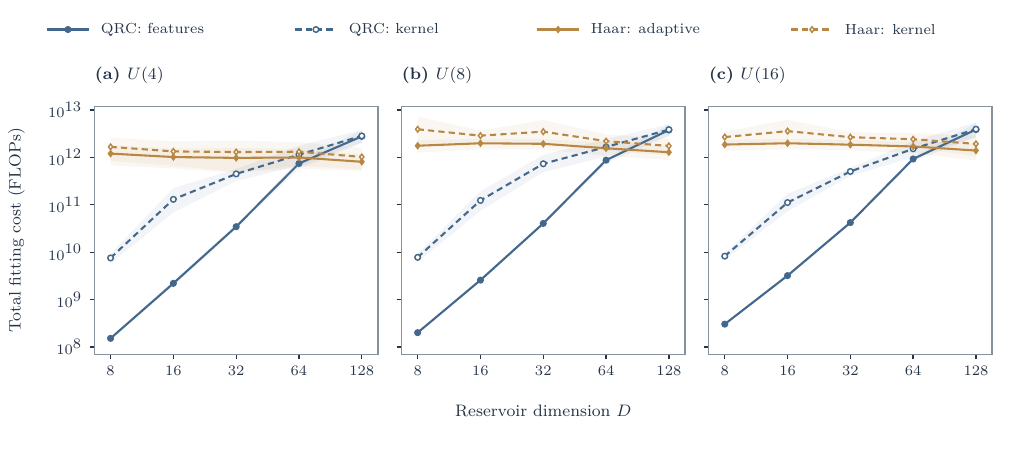}
\caption{Regression representation and total classical cost for the
three writer ensembles. Blue denotes finite QRC and ochre Haar.
Dashed curves use full sample-kernel eigendecomposition, as in
Fig.~\ref{fig:finite}(b); solid curves use the finite feature solver
and the Haar solver described above. All curves use the same targets,
selected penalties, and attained sample budgets, and include
representation construction and validation. Shading gives pointwise
95\% hierarchical-bootstrap intervals.}
\label{fig:solver_representation_control}
\end{figure*}

\begin{figure*}[t]
\centering
\includegraphics[width=\textwidth]{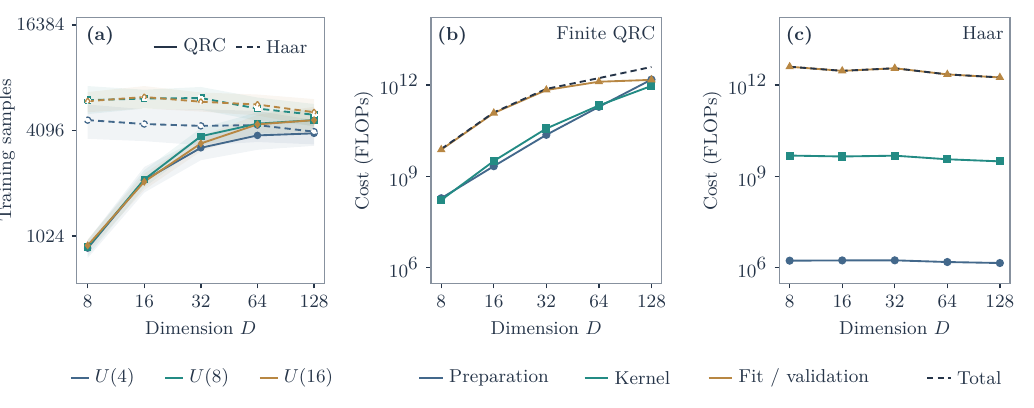}
\caption{Data requirements and cost decomposition for the full-kernel
comparison. (a) Mean attained training sizes: solid curves denote QRC,
dashed curves Haar, and colors the three writer families; bands give
pointwise 95\% intervals. (b,c) Costs for the $U(8)$ family only.
Here colors instead distinguish state/input preparation, kernel
construction, and fitting with validation; the dark dashed line is
their sum. These are the same learning runs as Fig.~\ref{fig:finite}(b),
with no additional fits.}
\label{smrw:samples-costs}
\end{figure*}

The plotted counts charge $2abc$ operations for a real
$(a\times b)(b\times c)$ product, $8abc$ for a complex product,
and $9n^3$ for symmetric eigendecomposition; lower-order
preprocessing and validation terms are also included. Costs cover one standalone fit from raw
histories and supplied labels, excluding device/label generation,
sample-budget search, test prediction, and I/O. They are estimates
for the specified algorithms, not measured runtimes.

An equivalent ridge fit can use the feature covariance. For an
$n\times p$ centered, normalized feature matrix $X$, $K=XX^{\mathsf T}$,
centered labels $y$, and $\lambda>0$,
\begin{equation}
 (X^{\mathsf T}X+n\lambda I_p)^{-1}X^{\mathsf T}y
 =X^{\mathsf T}(K+n\lambda I_n)^{-1}y.
 \label{smrw:primal-dual}
\end{equation}
After feature construction, the respective dense training costs are
\begin{equation}
 \begin{array}{lll}
 \text{feature covariance:}&p\times p,&O(np^2+p^3),\\
 \text{sample kernel:}&n\times n,&O(n^2p+n^3).
 \end{array}
 \label{smrw:primal-dual-cost}
\end{equation}
Thus $p\ll n$ favors feature-space fitting without changing the
predictor. A $D\times D$ density matrix has $D^2$ real Hermitian
coordinates and at most $D^2-1$ independent traceless directions.
State simulation and any exact coordinate compression must still
be counted.

Figure~\ref{fig:solver_representation_control} compares these solver
choices. The finite implementation retains $p=D^2$ coordinates and
chooses the smaller system. For the generic writers here, the explicit
Haar dictionary has $p=3^8-1=6560$ coordinates; its supplementary
solver switches from the recurrence to this dictionary at $n=p$.
This is the label \textit{Haar: adaptive}. The main curves always
use the recurrence and full sample-kernel decomposition.

The original learning runs used equivalent feature/kernel solvers,
including exact compression at large $D$. Their measured budgets
are reused for the main cost calculation; full-budget dense-kernel
fits were not rerun; numerical checks confirmed the equivalent fits.
The main cost model charges direct state propagation without
discounting these accelerations.

For the separate quantum readout benchmark in the Letter, a SWAP
outcome $x=\pm1$ has mean $\Tr(\rho\rho')$~\cite{SM-Ekert2002DirectEstimation}.
The raw-kernel estimate $\widehat k=D\overline x-1$ from $M$ independent
shots obeys
$\Pr(|\widehat k-k|>\epsilon)\le2\exp[-M\epsilon^2/(2D^2)]$.
Fresh preparation of two length-$L$ histories per shot therefore
uses $O(LD^2\epsilon^{-2}\log(2/\delta))$ reservoir updates at confidence
$1-\delta$. This count is specific to independent SWAP estimation.

\subsection{Target construction and learning protocol}

For each $d=2,4,8$, twelve Haar writers $w\in U(2d)$ are embedded as
$W_D=w\otimes I_{D/d}$ at $D=8,16,32,64,128$.
Four independent Haar routers per writer and dimension give 48 devices
per family and $D$. The matched recurrence uses the same writer's
actual local kernels, unchanged by the embedding.

We construct the target using the feature transfer
$F=\Phi_LT_D^{\rm raw}$ from Eq.~\eqref{smfc:metric}.
Let $v_D=\mathbb E_P\|F-\mathbb E_PF\|^2$ and
$v_H=\mathbb E_P\|\Phi_L-\mathbb E_P\Phi_L\|^2$ be the centered
feature energies. The normalized features used in Sec.~\ref{smgeom:section}
then satisfy
\begin{equation}
 \bar\phi_D=\bar\phi_HT_D,\qquad
 T_D=\sqrt{v_H/v_D}\,T_D^{\rm raw}.
 \label{smgeom:transfer}
\end{equation}
For the generic writers here, the local source and propagation kernels
have ranks two and three. Invertible local coordinate changes identify
their temporal features with the $3^L-1$ nonconstant products of
$1,h,z$, which remain independent after centering under the independent
uniform input law. The Haar coefficients $T_Dw$ are therefore unique
for the target $\bar\phi_Dw$. A unit top right singular vector of $T_D$
is orthogonal to its null space, so it also minimizes the finite weights
for its target and attains
\begin{equation}
 \Gamma_D=\|T_D\|_{\rm op}^2.
 \label{smgeom:exact-ratio}
\end{equation}
If the Haar dictionary is redundant, first project out coefficient
directions representing the zero function, orthogonally in the
Euclidean coefficient inner product, and use the projected transfer.
Each device's maximizing readout is centered and scaled to unit
population variance before generating learning data; this rescaling
does not change the weight ratio.
Population moments are evaluated using tensor products of
$1$, $(h-\pi/4)/\sqrt{2/3-\pi^2/16}$, and $\sqrt3z$.
The target changes with the device and $D$; maximizing the theoretical
weight ratio does not assert maximal empirical sample demand.

\begin{table}[htb]
\centering
\small
\caption{Learning protocol for Fig.~\ref{fig:finite}. Both models
receive paired inputs and noisy labels.}
\label{smrw:protocol}
\begin{tabular}{@{}p{.30\textwidth}p{.64\textwidth}@{}}
\toprule
Quantity & Setting \\
\midrule
Input histories & Independent length-eight histories, reset to $I_D/D$;
iid $z_t\sim\mathrm U[-1,1]$ and $u_t=(z_t+1)/2$ \\
Repetitions & Three independent data/noise repetitions per device \\
Training sizes & Powers of two from 16 to 8192; four Haar cases extended to 16384 \\
Validation/test sizes & 512 / 2048 histories \\
Label noise & Gaussian variance $0.1$ in training and validation; clean test targets \\
Ridge selection & 23 logarithmically spaced penalties in $[10^{-10},10]$, selected by validation MSE \\
Attainment criterion & First tested $n$ with mean clean test MSE $\le0.01$ over the three repetitions \\
\bottomrule
\end{tabular}
\end{table}

All devices attain the threshold; none is
discarded by performance. Test errors define the retrospective
attainment statistic, but never select penalties or targets.
Each device's cost is computed at its own attained $n$, before
averaging over devices. Figure~\ref{smrw:samples-costs} shows the
sample counts and cost components.

Pointwise 95\% intervals use 4000 hierarchical-bootstrap resamples,
first over writers and then routers, conditional on the three
data repetitions. Cost ratios use paired resampling. Input histories,
router draws, and label noise are paired across writer families.
The average $\Gamma_D$
is a mean of devicewise certificates, not a bound on the ratio of
ensemble-mean sample counts. This crossover is specific to the tested
ensemble and algorithms.

\section{Controlling memory and temporal interaction order}
\label{smdesign:methods}

\subsection{Kernel construction and memory controls}

The classical family in Eq.~\eqref{eq:design} sets
\begin{equation}
 a_0=R(1-\eta),\qquad C=R\eta I_2,\qquad
 B=\frac{1-R}{2}I_2,
 \quad 0<R<1,\quad 0\le\eta\le1.
 \label{smdesign:map}
\end{equation}
With $\boldsymbol s=(h,z)^{\mathsf T}$, the local feature maps are
$\sqrt{(1-R)/2}\,\boldsymbol s$ and
$\sqrt R\,(\sqrt{1-\eta},\sqrt\eta\,\boldsymbol s)$.
Their inner products define positive-semidefinite kernels, as does
the sum of products in Eq.~\eqref{eq:age}. Since
$\|\boldsymbol s\|=1$, $|a|\le R$ and $|b|\le(1-R)/2$;
the omitted source ages $\ell\ge L$ contribute at most $R^L/2$.
This proves uniform convergence of the infinite-history kernel,
without requiring a unitary realization.
For every input, $a(u,u)=R$: varying $\eta$ therefore preserves
diagonal retention while changing propagation overlaps between
distinct input histories. It does not fix the full history spectrum.

Source age $\ell$ carries a factor $R^\ell$, while $\eta$ allocates
weight between constant and input-dependent propagation. Trace
normalization removes the common source factor $(1-R)/2$, but
preserves these relative memory weights.

\subsection{Probing memory depth and interaction order}

Figure~\ref{fig:design} uses independent length-eight histories
with iid $z_j\sim\mathrm U[-1,1]$, $u_j=(z_j+1)/2$, and lag zero
current. The products below have zero population mean; their
prefactors give unit variance. These orthogonal probes follow
the information-processing-capacity approach~\cite{SM-Dambre2012}.

Panel (a) shifts a third-order product across six memory depths:
\begin{equation}
 f_\ell=3^{3/2}z_{\ell-2}z_{\ell-1}z_\ell,
 \qquad \ell=2,\ldots,7.
 \label{smdesign:memory-targets}
\end{equation}
We fix $\eta=1/4$ and evaluate
$R\in\{0.20,0.50,0.75,0.90,0.99\}$.

Panel (b) fixes the oldest required lag at seven and varies order:
\begin{equation}
 f_{\mathcal A}=3^{|\mathcal A|/2}\prod_{j\in\mathcal A}z_j,\qquad
 \mathcal A\subseteq\{0,\ldots,7\},\quad 7\in\mathcal A.
 \label{smdesign:order-targets}
\end{equation}
The $128$ targets are grouped by interaction order $k=|\mathcal A|$,
with $\binom7{k-1}$ targets at order $k$. At fixed $R=0.9$, we scan
$\eta\in\{0.05,0.25,0.40,0.55,0.70,0.85,0.99\}$.
The feature with an odd source at lag seven, the specified $k-1$
younger odd factors, and constants elsewhere has squared amplitude
proportional, at fixed $R$, to
\begin{equation}
 \eta^{k-1}(1-\eta)^{8-k}.
 \label{smdesign:order-weight}
\end{equation}
This is a dictionary weight, not a kernel eigenvalue: other features
need not be orthogonal, so this weight alone does not determine
finite-sample performance.

\subsection{Training and statistical evaluation}

Each of eight datasets has 512 training, 2048 validation, and 8192
test histories, initialized at zero kernel state. Training and
validation labels have independent Gaussian noise of variance $0.1$;
test labels are clean. Each target uses the same inputs and noise
across its parameter scan. We use the common regression protocol
of Sec.~\ref{smrw:section}. Each target and parameter point selects among 25
penalties $\lambda=10^{-10+j/2}$, $j=0,\ldots,24$, by validation
MSE. Test NMSE is clean test MSE divided by clean-target empirical
variance (divisor 8192); test labels enter neither fitting nor
selection. The plotted score is $1-\mathrm{NMSE}$, without clipping.

Panel (b) first averages all targets within each order and dataset;
both panels then average eight dataset scores. Pointwise 95\%
intervals have half-width $t_{0.975,7}s/\sqrt8$, where $s$ is their
sample standard deviation. These quantify dataset variability
conditional on the task bank, without simultaneous coverage.
All scan points and targets are retained. The grids and task bank
were chosen exploratorily, and the eight paired datasets reuse
exploration seeds; these are mechanism diagnostics, not an
independent confirmation set.

\section{Comparison with Hamiltonian reservoirs}
\label{sm:narma-fair}

\subsection{NARMA10 with continuous input streams}

Figure~\ref{fig:performance} compares the recursive kernel with two
Hamiltonian reservoirs on bounded-input NARMA10,
\begin{align}
 y_{t+1}&=0.3y_t+0.05y_t\sum_{j=0}^{9}y_{t-j}
              +1.5x_tx_{t-9}+0.1,\nonumber\\
 x_t&\stackrel{\mathrm{iid}}{\sim}\mathrm{Uniform}[0,0.2],
 \qquad u_t=5x_t.
 \label{smnarma:bounded-task}
\end{align}
The coefficients follow Ref.~\cite{SM-Paquot2012}; the narrower input
range also appears in Ref.~\cite{SM-Kobayashi2025}. After receiving
$u_t$, each model predicts $y_{t+1}$ without target feedback or
added noise. With the ten initial outputs set to zero, these inputs
keep $0\le y_t\le0.3$; no clipping or realization filtering is used.

Eight length-16384 streams share the zero-based, half-open splits
in Table~\ref{smnarma:splits}. Every model processes all inputs from
time zero, including washout and gaps, without resetting at split
boundaries.

\begin{table}[h]
\centering
\caption{Common endpoint intervals. Learning curves use nested training
suffixes; validation and test remain fixed.}
\label{smnarma:splits}
\begin{tabular}{lc}
\hline
Block & Interval \\
\hline
Initial washout & $[0,1024)$ \\
Training pool & $[1024,5120)$ \\
Selected training suffix & $[5120-n,5120)$, $n=256,512,1024$ \\
Validation & $[5632,7680)$ (2048 targets) \\
Test & $[8192,16384)$ (8192 targets) \\
\hline
\end{tabular}
\end{table}

For Eq.~\eqref{eq:design}, let $t,s$ count processed inputs, with
zero denoting the empty history rather than the array index above.
The uncentered kernel between complete prefixes obeys
\begin{align}
 \kappa_{t,s}
 &\equiv\kappa\bigl((u_1,\ldots,u_t),(u_1,\ldots,u_s)\bigr)
 =b(u_t,u_s)+a(u_t,u_s)\kappa_{t-1,s-1},\nonumber\\
 \kappa_{0,s}&=\kappa_{t,0}=0.
 \label{smnarma:prefix-kernel}
\end{align}
The recursion pairs equal-age memory components and stops after
$\min(t,s)$ steps; each diagonal entry retains its complete prefix.
All models evolve continuously through washout and gaps, with kernels
evaluated at the selected endpoints and no fixed input window or
truncation of physical states. The training size $n$ counts selected
endpoints; the size comparison uses $n=1024$.

\subsection{Hamiltonians and full-state readout}

The classical model uses the two-control family $(R,\eta)$ of
Sec.~\ref{smdesign:methods}, with equal even/odd source weights.
Its common source scale cancels under normalization. It has no
physical size, so the size comparison repeats identical fits.

The transverse-field Ising reservoir~\cite{SM-Fujii2017,SM-Martinez2020,SM-Martinez2021}
has controls $(\Delta t,h_z)$ and
\begin{equation}
 H_{\rm I}=\sum_{i<j}J_{ij}X_iX_j+h_z\sum_i Z_i,
 \quad U_{\rm I}=e^{-i\Delta tH_{\rm I}},
 \quad J_{ij}\sim\mathrm{Uniform}[-1/2,1/2].
 \label{smnarma:ising}
\end{equation}
The couplings retain this scale at every size.

For complex SYK, controls $(\Delta t,\alpha)$ mix quartic and
quadratic terms. With $P_p=c_jc_i$ for $p=(i,j)$, $i<j$,
\begin{align}
 H_4&=-4\sum_{p,q}J_{pq}P_p^\dagger P_q,
 &H_2&=\sum_{i,j}\kappa_{ij}c_i^\dagger c_j,\nonumber\\
 U_{\rm S}&=\exp[-i\Delta t(H_4+\alpha H_2)].
 \label{smnarma:syk}
\end{align}
Independent Hermitian Gaussian matrices $J,\kappa$ have zero mean
and variances $\mathbb E|J_{pq}|^2=N^{-3}$ and
$\mathbb E|\kappa_{ij}|^2=(2N)^{-1}$, including diagonals.
Evolution uses the full Fock space, retaining charge-sector
coherences. Both models use raw $\Delta t$ with the stated coupling
scales, without operator-norm normalization or trace shifting.

Here $N$ includes the reset input qubit, $D=2^{N-1}$ is the retained
dimension, and SYK has $N$ complex modes. Starting from $I_D/D$,
each step prepares the input in
$\sqrt{1-u_t}|0\rangle+\sqrt{u_t}|1\rangle$, applies the same joint
unitary, and traces out the input. These Hamiltonians generate the
complete joint evolution, without a separate Haar router.

Readout uses the full retained-state Hilbert--Schmidt kernel
\begin{equation}
 K_{st}=D\operatorname{Re}\Tr(\rho_s\rho_t).
 \label{smnarma:state-kernel}
\end{equation}
This gives ideal linear readout of all retained-state observables,
without shot noise or time multiplexing.

\subsection{Parameter selection and evaluation}

Each family has two controls and 64 fixed candidates, frozen before
their respective bounded-input fits. The candidate sets combine
independent preliminary screening with global, local, and reference
points; they are not uniform grids. The search coordinates and
intervals are
\begin{align}
 \log(1-R)&\in[\log0.01,\log0.8],
 &\log\eta&\in[\log10^{-4},\log0.99],\nonumber\\
 \log\Delta t_{\rm I}&\in[\log0.05,\log20],
 &\log h_z&\in[\log0.01,\log5],\nonumber\\
 \log\Delta t_{\rm S}&\in[\log0.001,\log100],
 &\log(1+\alpha)&\in[0,\log101].
 \label{smnarma:search-domain}
\end{align}
Equal control and candidate counts define the comparison budget.

Each repetition pairs a shared input stream with an independent
device per family and size. Couplings stay fixed across its controls;
devices are not selected. At every size and training budget, each
repetition selects controls and ridge anew using validation alone.
We use the common ridge-regression convention of
Sec.~\ref{smrw:section}, with 29 penalties
$\lambda=10^{-14},10^{-13.5},\ldots,1$. Full SVD or Gram
eigendecomposition implements the readout without rank truncation.

Selected models are fixed before test scoring. NMSE divides test
MSE by empirical test-target variance. Means use all eight paired
repetitions; pointwise 95\% percentile-bootstrap intervals use
50000 resamples of whole input--device pairs, identically resampled
across families, sizes, and budgets. This is an exploratory
comparison conditional on the fixed candidate sets and ideal,
noise-free readout.

%

\clearpage
\end{document}